\documentclass[5p,times,twocolumn]{elsarticle}        % Final version format
\usepackage{preamble_paper}
\usepackage{newcommands_paper}

\begin{document}

\title{Derivation of a Local Volume-Averaged Model and a Stable Numerical Algorithm for Multi-Dimensional Simulations of Conversion Batteries}

%%% Author list for elsarticle
\author[dlr,hiu,uu]{Tobias Schmitt}
\author[dlr,hiu,uu]{Arnulf Latz}
\author[dlr,hiu,uu]{Birger Horstmann\corref{cor1}}\ead[url]{birger.horstmann@dlr.de}

\cortext[cor1]{Corresponding author}
\address[dlr]{German Aerospace Center, Pfaffenwaldring 38-40, 70569 Stuttgart, Germany}
\address[hiu]{Helmholtz Institute Ulm, Helmholtzstra\ss{}e 11, 89069 Ulm, Germany}
\address[uu]{Ulm University, Institute of Electrochemistry, Albert-Einstein-Allee 47, 89069 Ulm, Germany}

\begin{abstract}	
In this article, we derive a general form of local volume-averaging theory and apply it to a model of zinc-air conversion batteries. Volume-averaging techniques are frequently used for the macroscopic description of micro-porous electrodes. We extend the existing method by including reactions between different phases and time-dependent volume fractions of the solid phases as these are continuously dissolved and reconstructed during operation of conversion batteries. We find that the constraint of incompressibility for multi-component fluids causes numerical instabilities in simulations of zinc-air battery cells. Therefore, we develop a stable sequential semi-implicit algorithm which converges against the fully implicit solution. Our method reduces the coupling of the variables by splitting the system of equations and introducing an additional iteration step. 
\end{abstract}

\begin{keyword}
	local volume-averaging theory \sep
	multi-component incompressibility constraint \sep
	sequential semi-implicit solver \sep
	numerical stability analysis \sep
	zinc-air battery
\end{keyword}

%	%%%%% Title page %%%%%%
\maketitle
%	\tableofcontents
%	%%%%%%%%%%%%%%%%%

%\todo[inline]{Dots for vector products}

\section{Introduction}
\label{sec:intro}

Next-generation batteries promise large energy densities while relying on cheap, safe, and ecologically friendly materials. Zinc-air batteries are a prominent example. However, there are some hurdles to overcome before rechargeable zinc-air batteries will become a widespread commercial reality. We contribute to this research by multi-dimensional cell-level simulations.

Electrodes and separators of zinc-air batteries are highly porous composites. Their pore sizes are much smaller than their cell sizes. The conversion of electrode structure takes place on an even smaller scale. Therefore, we cannot resolve the whole cell micro-structure on a single computational grid. Two approaches are feasible, either we perform detailed simulations of micro-scale processes or we perform mean-field simulations for the whole cell. The latter approach requires a theory of averaging transport and reaction equations. In this article, we discuss the local volume-averaging theory (LVA), formulate a model for averaged quantities, such as a mean concentration and mean potential, and develop a numeric integration algorithm.  

Our theory of local volume-averaging is based on the works of Chen \cite{Zhangxin1994,Chen1995} and  Whitaker \cite{Whitaker1967,Whitaker1984,Whitaker1999}. They assume a constant porosity and neglect reactions between phases. Thus, the solid volume fraction depends on space only. We derive a more general theory for space- and time-dependent volume fractions. Convection of liquid and solid phases as well as reactions at phase boundaries contribute to continued material redistribution. As reactions on phase boundaries, we consider the transport through interfaces and the conversion of phases.

Clark et al. \cite{Clark2018} review the current state of zinc-air battery modeling \cite{Schmitt2019,Stamm2017,Clark2019,Clark2017,Schroder2014_2,Schroder2014_1,Deiss2002,Isaacson1990,Mao1992,Sunu1978}. In our previous works on metal-air batteries \cite{Horstmann2013,Schmitt2019,Stamm2017,Clark2019,Clark2017,Danner2014}, we show how to incorporate convection by a multi-component incompressibility constraint for concentrated solutions (see \cref{eq:lva:ex:gce:lva_mcic}). Our definition differs from the standard incompressibility condition in computational fluid dynamics. Multi-dimensional simulations based on our transport model are numerically unstable because all variables are strongly coupled by the constraint of incompressibility. Our one-dimensional simulations, in contrast, are stable \cite{Stamm2017,Hoffmann2018}. Conventional approaches for the stabilization of the solver algorithm \cite{Patankar1980,Ferziger2002} cannot resolve this issue here. In this article, we develop a stable particle-number-conserving integration algorithm (see Sec. \ref{sec:lva}) and demonstrate multi-dimensional simulations of conversion type batteries (see Sec. \ref{sec:num}).

\section{Local Volume-Averaging Theory}
\label{sec:lva}
\begin{figure}[!tb]
	\centering
	\includegraphics[width=0.9\columnwidth]{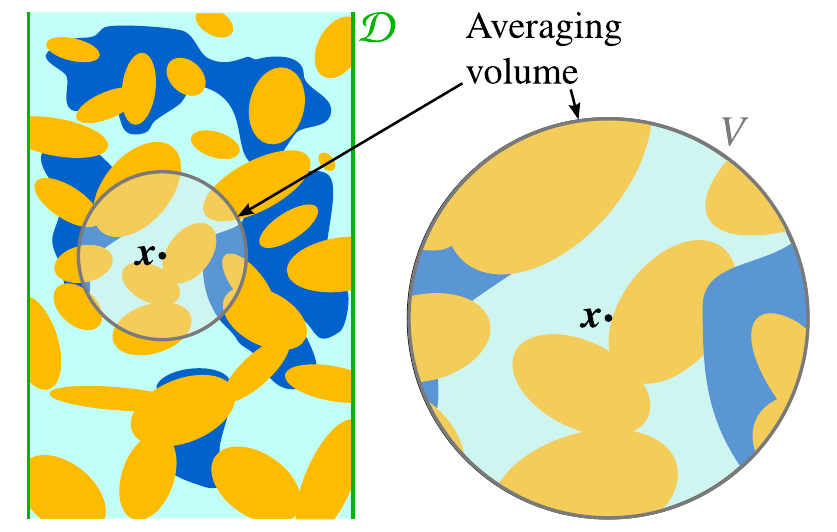}
	\caption{Volume-averaging over a multi-phase domain with a symmetric well-like weighting function $m(\myvec{x})$ represented by the averaging volume $V$. An average at position $\myvec{x}$ is obtained by an integration over $\mathcal{D}$ weighted by $m(\myvec{x}')$.}
	\label{fig:lva:ansatz}
\end{figure}

Local volume-averaging enables simulations of porous media on length scales, which are much larger than the typical grain size. The transport equations are stated with averaged, i.e., effective, field variables which are determined by an average over the local neighborhood. Figure \ref{fig:lva:ansatz} illustrates this principle. This approach is used phenomenologically in battery modelling, but a rigorous derivation and consistent formulation for conversion-type batteries is still missing. 
In this section, we derive a volume-averaging theory based on Refs. \cite{Zhangxin1994,Chen1995,Whitaker1967,Whitaker1984,Whitaker1999}, which includes space- and time-dependent volume fractions of all phases. Additionally, we incorporate convection and phase changing bulk and surface reactions. In \cref{div}, we introduce the concepts of local volume-averaging before applying them to the set of transport equations in \cref{sec:lva:ex}. Modeling choices are necessary to evaluate the resulting transport and reaction equations (see \cref{sec:lva:mod}).

\subsection{Definitions and Theory} \label{sec:lva:def}
\begin{figure}[!tb]
	\centering
	\includegraphics[width=1\columnwidth]{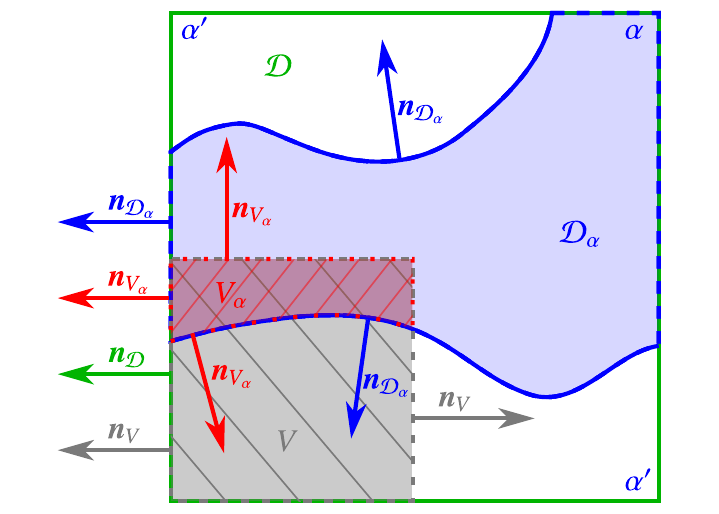}
	\vspace*{-7mm}
	\caption{Illustration of a multi-phase domain in the local volume-averaging theory. The averaging volume $V$ (gray hatched) is embedded in the global domain $\mathcal{D}$ (green). $V_{\alpha}$ and $\mathcal{D}_{\alpha}$ indicate the restrictions to phase $\alpha$ (red hatched and blue).}
	\label{fig:lva:surfaceNormalVectors}
\end{figure}

Let us define a region in space $\mathcal{D} \subseteq \mathbb{R}^{3}$ with the coordinate vector $\myvec{x} \in \mathcal{D}$ and let further $\Psi(t,\myvec{x})$ be a tensor field of rank $n$ defined on $\mathcal{D}$. We assume that $\mathcal{D}$ contains a number of phases, denoted with the index $\alpha$ (see \cref{fig:lva:ansatz}). Typically, these phases are solid, liquid, and gas. Let further $\mathcal{D}_{\alpha}$ be the part of $\mathcal{D}$, occupied by phase $\alpha$. The local volume-average of $\Psi$ is obtained by a spatial convolution of $\Psi(t,\myvec{x})$ with a weighting function $m(\myvec{x})$ \cite{Chen1995}. This function can be used to match the theory with the characteristics of experimental measuring devices \cite{Cushman1984,Baveye1984}. The weighting function must be strictly non-negative, normalized, and monotonically decreasing steeper than $1 / \abs{\myvec{x}}$. This implies that 
\begin{align}
	\int\limits_{\mathbb{R}^{3}} m(\myvec{x}) \mathrm{d}^{3}\myvec{x} = 1 \, .
	\label{eq:lva:weighting_function}
\end{align}
The spatial moments of order $n$ of the function $m(\myvec{x})$ are defined by $\int_{\mathbb{R}^{3}} \myvec{x}^{n} m(\myvec{x}) \mathrm{d}^{3}\myvec{x}$. It is required that all odd spatial moments vanish \cite{Quintard1994,Goltz1987}. We define the local volume-average as
\begin{align}
	\lvaSup{\Psi_{\alpha}} &\equiv \lvaSup{\Psi_{\alpha}} (t,\myvec{x})   \nonumber\\
	~
	&= \int\limits_{\mathcal{D}} \Psi_{\alpha}(t,\myvec{x}') \chi_{\alpha}(t,\myvec{x}') m(\myvec{x} - \myvec{x}') \mathrm{d}^{3}\myvec{x}'  \nonumber\\
	~
	&= \int\limits_{\mathbb{R}^{3}} \Psi_{\alpha}(t,\myvec{x} - \myvec{x}') \chi_{\alpha}(t,\myvec{x} - \myvec{x}') m(\myvec{x}') \mathrm{d}^{3}\myvec{x}' \, ,
	\label{eq:lva:lva}
\end{align}
with the index $\Psi_\alpha$ referring to $\Psi$ in phase $\alpha$. $\lvaSup{\Psi_{\alpha}}$ is the convolution product of $\Psi_{\alpha}\chi_{\alpha}$ with the weighting function $m$. The function $\chi_{\alpha}(t, \myvec{x})$ is the characteristic phase function of phase $\alpha$ (see \cref{fig:lva:surfaceNormalVectors}). It is defined by
\begin{align}
	\chi_{\alpha}(t, \myvec{x}) &=
	\begin{cases}
		1 \text{, if } \myvec{x} \text{ at time } t \text{ is in phase }\alpha  \\
		0 \text{, else } \\
	\end{cases} . \label{eq:lva:characteristic_phase_function}
\end{align}
We apply the local volume-averaging theory for the equations of fluid dynamics. In the following subsections Secs. \ref{div} and \ref{dt}, we discuss two especially relevant mathematical terms: the divergence of a vector field and the partial time derivative of a scalar field. Then we present our choice of the weighting function in Sec. \ref{weighting}. In Sec. \ref{averages}, we introduce intrinsic and surface averages needed for the formulation of battery models.

\subsubsection{Divergence of a Vector Field} \label{div}
$\lvaSup{\bn \myvec{\Psi}_{\alpha}}$ is evaluated as
\begin{align}
	\lvaSup{\bn_{\myvec{x}} \myvec{\Psi}_{\alpha}(t,\myvec{x})} &= \nonumber\\
	~
	&\hspace{-1.2cm} = \int\limits_{\mathcal{D}} \left[ \bn_{\myvec{x}'} \myvec{\Psi}_{\alpha}(t,\myvec{x}') \right] \chi_{\alpha}(t,\myvec{x}') m(\myvec{x} - \myvec{x}') \mathrm{d}^{3}\myvec{x}'  \label{eq:lva:lva_divergence_vector_proof1}\\
	~
	&\hspace{-1.2cm} = \phantom{-} \int\limits_{\mathcal{D}} \bn_{\myvec{x}'} \left[ \myvec{\Psi}_{\alpha}(t,\myvec{x}') \chi_{\alpha}(t,\myvec{x}') m(\myvec{x} - \myvec{x}') \right] \mathrm{d}^{3}\myvec{x}' \nonumber\\ 
	&\hspace{-0.83cm} - \int\limits_{\mathcal{D}} \myvec{\Psi}_{\alpha}(t,\myvec{x}') \left[ \bn_{\myvec{x}'} \chi_{\alpha}(t,\myvec{x}') \right] m(\myvec{x} -  \myvec{x}') \mathrm{d}^{3}\myvec{x}' \nonumber \\
	&\hspace{-0.83cm} - \int\limits_{\mathcal{D}} \myvec{\Psi}_{\alpha}(t,\myvec{x}') \chi_{\alpha}(t,\myvec{x}') \left[ \bn_{\myvec{x}'} m(\myvec{x} - \myvec{x}') \right] \mathrm{d}^{3}\myvec{x}'  \, .
	\label{eq:lva:lva_divergence_vector_proof2}
\end{align}
We transform the first term of \cref{eq:lva:lva_divergence_vector_proof2} with the divergence theorem into a surface integral over the boundary $\partial \mathcal{D}$ with the outward pointing normal vector $\myvec{n}_{\mathcal{D}}$. The characteristic phase function $\chi_{\alpha}(t,\myvec{x}')$ is a Heaviside-like function. Its derivative is a delta distribution times the surface normal vector \cite{heaviside}
\begin{align}
	\bn_{\myvec{x}'} \chi_{\alpha}(t,\myvec{x}') = \delta\left( d_{\partial \mathcal{D}_{\alpha} \backslash \partial\mathcal{D}} \right) \myvec{n}_{\alpha', \alpha}  \, , \label{eq:lva:grad_characteristic_phase_function}
\end{align}
with the distance function $d_{\partial \mathcal{D}_{\alpha} \backslash \partial\mathcal{D}}$ to the boundary $\partial \mathcal{D}_{\alpha} \backslash \partial\mathcal{D}$ ($\partial \mathcal{D}_{\alpha}$ without $\partial \mathcal{D}$) and the surface normal vector $\myvec{n}_{\alpha', \alpha}$ pointing from phase $\alpha'$ to $\alpha$. The surface $\partial \mathcal{D}_{\alpha} \backslash \partial\mathcal{D}$ is the boundary between the phases $\alpha$ and $\alpha'$. The domain boundary $\partial \mathcal{D}$ is by definition no phase boundary and thus  ${\left.\bn\chi_{\alpha}(\myvec{x})\right|_{\myvec{x} \in \partial\mathcal{D}} = 0}$ (see \cref{fig:lva:surfaceNormalVectors}). This reduces the volume integral in term two of \cref{eq:lva:lva_divergence_vector_proof2} into a surface integral. To transform term three, we use
\begin{align}
	\bn_{\myvec{x}'} m(\myvec{x} - \myvec{x}') \mathrm{d}^{3}\myvec{x}' = - \bn_{\myvec{x}} m(\myvec{x} - \myvec{x}') \mathrm{d}^{3}\myvec{x}'  \, .
	\label{eq:lva:transformGradWeightFunc}
\end{align}
Writing a single surface integral for the whole surface ($\alpha'$ can change within the integration boundaries), we get
\begin{align}
	\lvaSup{\bn_{\myvec{x}} \myvec{\Psi}_{\alpha}(t,\myvec{x})} &=  \nonumber\\
	~
	&\hspace{-1.5cm} = \phantom{+} \oint\limits_{\partial \mathcal{D}}  \myvec{\Psi}_{\alpha}(t,\myvec{x}') \myvec{n}_{\mathcal{D}}(t,\myvec{x}') \chi_{\alpha}(t,\myvec{x}') m(\myvec{x} - \myvec{x}') \mathrm{d}^{2}\myvec{x}'  \nonumber\\
	&\hspace{-1.22cm} - \int\limits_{\partial \mathcal{D}_{\alpha} \backslash \partial\mathcal{D}} \myvec{\Psi}_{\alpha}(t,\myvec{x}') \myvec{n}_{\alpha', \alpha}(t,\myvec{x}') m(\myvec{x} -  \myvec{x}') \mathrm{d}^{2}\myvec{x}'  \nonumber\\
	&\hspace{-1.22cm} + \bn_{\myvec{x}}  \int\limits_{\mathcal{D}} \myvec{\Psi}_{\alpha}(t,\myvec{x}') \chi_{\alpha}(t,\myvec{x}') m(\myvec{x} - \myvec{x}') \mathrm{d}^{3}\myvec{x}'  \, .
	\label{eq:lva:lva_divergence_vector_proof3}
\end{align}
Using the definition of $\chi_{\alpha}$ reduces the integration boundary of the first term to $\partial\mathcal{D}_{\alpha}$ on $\partial\mathcal{D}$, denoted by ${\partial\mathcal{D}_{\alpha} \!\cap\! \partial\mathcal{D}}$. On this surface, it holds ${\myvec{n}_{\mathcal{D}} = \myvec{n}_{\mathcal{D}_{\alpha}}}$. In the second term, we use that ${\myvec{n}_{\alpha', \alpha} = -\myvec{n}_{\alpha, \alpha'} = \myvec{n}_{\mathcal{D}_{\alpha}}}$ holds on ${\partial\mathcal{D}_{\alpha} \backslash \partial\mathcal{D}}$. We simplify the last term with \cref{eq:lva:lva}. Thus, we get
\begin{align}
	\lvaSup{\bn_{\myvec{x}} \myvec{\Psi}_{\alpha}(t,\myvec{x})} =&   \int\limits_{\partial\mathcal{D}_{\alpha} \cap \partial\mathcal{D}}  \myvec{\Psi}_{\alpha}(t,\myvec{x}') \myvec{n}_{\mathcal{D}}(t,\myvec{x}') m(\myvec{x} - \myvec{x}') \mathrm{d}^{2}\myvec{x}'  \nonumber\\
	&+ \int\limits_{\partial \mathcal{D}_{\alpha} \backslash \partial\mathcal{D}} \myvec{\Psi}_{\alpha}(t,\myvec{x}') \myvec{n}_{\mathcal{D}_{\alpha}}(t,\myvec{x}') m(\myvec{x} -  \myvec{x}') \mathrm{d}^{2}\myvec{x}'  \nonumber\\
	&+ \bn_{\myvec{x}} \lvaSup{\myvec{\Psi}_{\alpha}(t,\myvec{x})}  \, .
	\label{eq:lva:lva_divergence_vector_proof4}
\end{align}
We see that both surface integrals are complementing each other to a closed surface integral over $\partial\mathcal{D}_{\alpha}$. This results in
\begin{align}
	\lvaSup{\bn_{\myvec{x}} \myvec{\Psi}_{\alpha}(t,\myvec{x})} &=  \bn_{\myvec{x}} \lvaSup{\myvec{\Psi}_{\alpha}(t,\myvec{x})}  \nonumber\\
	~
	&\hspace{-1.2cm}+ \oint\limits_{\partial \mathcal{D}_{\alpha}} \myvec{\Psi}_{\alpha}(t,\myvec{x}') \myvec{n}_{\mathcal{D}_{\alpha}}(t,\myvec{x}')  m(\myvec{x} -  \myvec{x}') \mathrm{d}^{2}\myvec{x}'   \, .
	\label{eq:lva:lva_divergence_vector}
\end{align}

\subsubsection{Time Derivative of a Scalar Field}\label{dt}
Next, we evaluate the average of a partial time derivative $\lvaSup{\partial_{t} \Psi_{\alpha}}$. As a preparation, we express the continuity equation for $\chi_{\alpha}$ in an Eulerian form with a production term $Q_{\chi_{\alpha}}$ on the right hand side,
\begin{align}
\partial_{t} \chi_{\alpha}(t,\myvec{x}') + \velSurf{c}{\alpha, \alpha'}(t,\myvec{x}') \cdot \bn \chi_{\alpha}(t,\myvec{x}') = Q_{\chi_{\alpha}}(t,\myvec{x}')  \, , \label{eq:lva:lva_continuity_chi1}
\end{align}
with the phase field convection velocity $\velSurf{c}{\alpha, \alpha'}$. We express $Q_{\chi_{\alpha}}$ by a volume creation rate per surface area $\velSurf{$\ast$}{\alpha, \alpha'}$, which we call growth velocity. The phase modifying reactions take place only on a phase surface. Thus, we set the production term to 
\begin{align}
	Q_{\chi_{\alpha}}(t,\myvec{x}') &= \left(\velSurf{$\ast$}{\alpha, \alpha'} \cdot \myvec{n}_{\alpha, \alpha'}\right) \! (t,\myvec{x}') \,  \delta\left( d_{\partial \mathcal{D}_{\alpha} \backslash \partial\mathcal{D}} \right)  \nonumber\\
	~
	&\hspace*{-4.2mm} \stackrel{\textrm{\cref{eq:lva:grad_characteristic_phase_function}}}{=} - \velSurf{$\ast$}{\alpha, \alpha'}(t,\myvec{x}') \cdot \bn \chi_{\alpha}(t,\myvec{x}')   \, .
	\label{eq:lva:lva_production_term_chi}
\end{align}
Inserting this in \cref{eq:lva:lva_continuity_chi1}, we get
\begin{align}
	\partial_{t} \chi_{\alpha}(t,\myvec{x}') = - \left( \velSurf{c}{\alpha, \alpha'} + \velSurf{$\ast$}{\alpha, \alpha'} \right)\!(t,\myvec{x}') \cdot \bn \chi_{\alpha}(t,\myvec{x}')  \, ,  \label{eq:lva:lva_continuity_chi}
\end{align}
with the total phase field velocity $\velSurf{tot}{\alpha, \alpha'} = \velSurf{c}{\alpha, \alpha'} + \velSurf{$\ast$}{\alpha, \alpha'}$. Now, we can derive the volume-average of $\partial_{t}\Psi_{\alpha}$. We start with
\begin{align}
	\partial_{t} \lvaSup{\Psi_{\alpha}(t,\myvec{x})} = &\hspace{0.35cm} \int\limits_{\mathcal{D}} \left[ \partial_{t} \Psi_{\alpha}(t,\myvec{x}') \right] \chi_{\alpha}(t,\myvec{x}') m(\myvec{x} - \myvec{x}') \mathrm{d}^{3}\myvec{x}'  \nonumber\\
	&+ \int\limits_{\mathcal{D}}  \Psi_{\alpha}(t,\myvec{x}') \left[ \partial_{t} \chi_{\alpha}(t,\myvec{x}') \right] m(\myvec{x} - \myvec{x}') \mathrm{d}^{3}\myvec{x}'  \nonumber\\
	&+ \int\limits_{\mathcal{D}}  \Psi_{\alpha}(t,\myvec{x}') \chi_{\alpha}(t,\myvec{x}') \left[ \partial_{t} m(\myvec{x} - \myvec{x}') \right] \mathrm{d}^{3}\myvec{x}'  \, . \label{eq:lva:lva_time_derivative_proof1}
\end{align}
The first term of \cref{eq:lva:lva_time_derivative_proof1} is identified with $\lvaSup{\partial_{t}\Psi_{\alpha}}$. We insert \cref{eq:lva:grad_characteristic_phase_function,eq:lva:lva_continuity_chi} into the second term and drop the last one because $m(x)$ is time independent. This transforms \cref{eq:lva:lva_time_derivative_proof1} into
\begin{align}
	\partial_{t} \lvaSup{\Psi_{\alpha}(t,\myvec{x})} &= \lvaSup{\partial_{t} \Psi_{\alpha}(t,\myvec{x})}  \nonumber\\
	&\hspace{-1.5cm} + \int\limits_{\mathcal{D}} \Psi_{\alpha}(t,\myvec{x}') \left[ -\velSurf{tot}{\alpha, \alpha'} \cdot \bn \chi_{\alpha} \right]\!\! (t,\myvec{x}') \, m(\myvec{x} - \myvec{x}') \mathrm{d}^{3}\myvec{x}'  \nonumber\\  
	~
	&= \lvaSup{\partial_{t} \Psi_{\alpha}(t,\myvec{x})}  \nonumber\\
	&\hspace{-1.5cm} + \int\limits_{\partial \mathcal{D}_{\alpha} \backslash \partial\mathcal{D}} \Psi_{\alpha}(t,\myvec{x}') \left[ \velSurf{tot}{\alpha, \alpha'} \cdot \myvec{n}_{\alpha, \alpha'} \right] \!\!(t,\myvec{x}') \, m(\myvec{x} - \myvec{x}') \mathrm{d}^{2}\myvec{x}'  \, .
	\label{eq:lva:lva_time_derivative_proof}
\end{align}
Reshuffling the terms yields
\begin{align}
	\lvaSup{\partial_t \Psi_{\alpha}(t,\myvec{x})} &= \partial_{t} \lvaSup{\Psi_{\alpha}(t,\myvec{x})}  \nonumber\\
	&\hspace{-2cm}- \int\limits_{\partial \mathcal{D}_{\alpha} \backslash \partial\mathcal{D}} \Psi_{\alpha}(t,\myvec{x}') \left[ \velSurf{tot}{\alpha, \alpha'} \cdot \myvec{n}_{\alpha, \alpha'} \right] \!\!(t,\myvec{x}')   \, m(\myvec{x} - \myvec{x}') \mathrm{d}^{2}\myvec{x}'  \, .
	\label{eq:lva:lva_time_derivative}
\end{align}

\subsubsection{Choice of the Weighting Function}\label{weighting}
A possible choice of the weighting function $m$ is a symmetric well-like function as suggested in Refs. \cite{Chen1995,Whitaker1967,Whitaker1984,Slattery1967}. This implies
\begin{align}
	m(\myvec{x}) = \frac{1}{\abs{V}} \chi_{V}(\myvec{x}) &=
	\begin{cases}
		\frac{1}{\abs{V}}   &   \text{, if } \myvec{x} \text{ is in } V \\
		0   &   \text{, else } \\
	\end{cases}
	\, ,
	\label{eq:lva:sphericall_well_weighting_function}
\end{align}
with the characteristic function $\chi_{V}$ for the volume $V$ of size $\abs{V}$. We denote with $V_{\alpha}$ the restriction of $V$ to phase $\alpha$. Examples of spherically symmetric and rectangular weighting functions are shown in \cref{fig:lva:ansatz,fig:lva:surfaceNormalVectors}. Inserting \cref{eq:lva:sphericall_well_weighting_function} into \cref{eq:lva:lva,eq:lva:lva_divergence_vector,eq:lva:lva_time_derivative} reduces the integral boundaries (see \cref{fig:lva:surfaceNormalVectors}).
\begin{align}
	&\lvaSup{\Psi_{\alpha}} = \frac{1}{\abs{V}} \int\limits_{V_{\alpha}} \Psi_{\alpha}(t,\myvec{x} - \myvec{x}') \mathrm{d}^{3}\myvec{x}'  \, , \label{eq:lva:lva_short}\\
	~
	&\lvaSup{\bn \myvec{\Psi}_{\alpha}} = \bn \lvaSup{\myvec{\Psi}_{\alpha}} + \frac{1}{\abs{V}} \int\limits_{ V_ \cap \partial \mathcal{D}_{\alpha} \hspace*{-0.7cm} } \myvec{\Psi}_{\alpha}(t,\myvec{x} - \myvec{x}') \myvec{n}_{V_{\alpha}}(t,\myvec{x} - \myvec{x}') \mathrm{d}^{2}\myvec{x}'  \, , \label{eq:lva:lva_divergence_vector_short}\\
	~
	&\lvaSup{\partial_t \Psi_{\alpha}} = \partial_{t} \lvaSup{\Psi_{\alpha}}
	\nonumber\\
	&- \frac{1}{\abs{V}} \int\limits_{ V \cap (\partial \mathcal{D}_{\alpha} \backslash \partial \mathcal{D}) \hspace*{-1.2cm} } \Psi_{\alpha}(t,\myvec{x} - \myvec{x}') \left( \velSurf{tot}{\alpha, \alpha'} \cdot \myvec{n}_{\alpha, \alpha'} \right) (t,\myvec{x} - \myvec{x}') \mathrm{d}^{2}\myvec{x}'   \, . \label{eq:lva:lva_time_derivative_short}
\end{align}

\subsubsection{Intrinsic and Surface Averages} \label{averages}
In battery models, the intrinsic volume average $\lvaIn{\Psi}$ appears. We define it analogous to \cref{eq:lva:lva_short} as the average of $\Psi$ within a specific phase,
\begin{align}
	\lvaIn{\Psi}_{\alpha} = \frac{1}{\abs{V_{\alpha}}} \int\limits_{V_{\alpha}} \Psi_{\alpha}(t,\myvec{x} - \myvec{x}') \mathrm{d}^{3}\myvec{x}'  \, ,
	\label{eq:lva:intrinsic_lva_short}
\end{align}
The volume fraction of phase $\alpha$ in the averaging-volume $V$ is ${\epsA(t, \myvec{x}) = \abs{V_{\alpha}}(t, \myvec{x}) / \abs{V}}$. Thus, the two volume-averages (\cref{eq:lva:lva_short,eq:lva:intrinsic_lva_short}) are related by	
\begin{equation}
\lvaSup{\Psi_{\alpha}} = \epsA \lvaIn{\Psi}_{\alpha}. 
\end{equation}
Analogous to the local volume-average, we define the local surface-average over the whole phase surface $\lvaSurfC{\Psi}_{\alpha}$ and over individual surface segments $\lvaSurfP{\Psi}_{\alpha, \alpha'}$,
\begin{align}
	&\lvaSurfC{\Psi}_{\alpha} = \frac{1}{\abs{A_{\alpha}}} \int\limits_{A_{\alpha}} \Psi_{\alpha}(t,\myvec{x} - \myvec{x}') \mathrm{d}^{2}\myvec{x}'  \, , \label{eq:lva:lsa_whole_short} \\
	~
	&\lvaSurfP{\Psi}_{\alpha, \alpha'} = \frac{1}{|A_{\alpha, \alpha'}|} \int\limits_{A_{\alpha, \alpha'}} \Psi_{\alpha}(t,\myvec{x} - \myvec{x}') \mathrm{d}^{2}\myvec{x}'  \, . \label{eq:lva:lsa_part_short}
\end{align}
We now define the surface areas $A_{\alpha}$ and $A_{\alpha, \alpha'}$ by (compare with \cref{fig:lva:surfaceNormalVectors})
\begin{align}
	&A_{\alpha} = V \cap \partial\mathcal{D}_{\alpha} \, , \nonumber\\
	~
	&A_{\alpha, \alpha'} = 
		\begin{cases}
		V \cap \partial\mathcal{D}_{\alpha'}   &   \text{, if } \alpha' \text{ is related to a phase} \\
		 V \cap \partial\mathcal{D}             &   \text{, if } \alpha' \text{ is a domain boundary}\\
	\end{cases}   \, .
	\label{eq:lva:lsa_surfAreas}
\end{align}
For simplicity, we treat the domain boundary as an additional phase in \cref{eq:lva:lsa_surfAreas}. 

The surface integral of the intrinsic mean $\lvaIn{\Psi}_\alpha$ over a phase boundary appears in the derivation of a volume-averaged battery model in \cref{sec:lva:ex}. As a preliminary, we represent the field variable $\Psi_{\alpha}$ as the sum of its intrinsic mean and a variation field $\lvaVar{\Psi}_{\alpha}$ (see Ref. \cite[chapter 1.2]{Whitaker1999}). 
\begin{align}
	\Psi_{\alpha}(t, \myvec{x}) = \lvaIn{\Psi}_{\alpha}(t, \myvec{x}) + \lvaVar{\Psi}_{\alpha}(t, \myvec{x})  \, .
	\label{eq:lva:spatial_decomp_psi}
\end{align}
The variation field is the deviation of the intrinsic mean from $\Psi_{\alpha}$. In \cref{eq:lva:spatial_decomp_psi}, $\lvaIn{\Psi}_{\alpha}$ is defined at the center of the averaging volume. We use a Taylor expansion to obtain $\lvaIn{\Psi}_{\alpha}$ on the phase boundary
\begin{align}
	\lvaIn{\Psi}_{\alpha}(t, \myvec{x} + \myvec{x}') =& \, \lvaIn{\Psi}_{\alpha}(t, \myvec{x}) + \left( \myvec{x}' \cdot \bn_{\myvec{x}} \right) \lvaIn{\Psi}_{\alpha}(t, \myvec{x})  \nonumber\\
	&+ \frac{1}{2} \left( \myvec{x}' \otimes \myvec{x}' : \bn_{\myvec{x}} \otimes \bn_{\myvec{x}} \right) \lvaIn{\Psi}_{\alpha}(t, \myvec{x}) + \dots \, ,
	\label{eq:lva:taylor_psi}
\end{align}
where $\otimes$ is the dyadic product and $:$ marks the complete contraction of the left and right matrices. The vector $\myvec{x}'$ points from the center of the averaged volume to a point on $\mathcal{D}_{\alpha}$. With \cref{eq:lva:taylor_psi} we evaluate the surface integral of $\lvaIn{\Psi}_\alpha$ as
\begin{align}
	&\int\limits_{A_{\alpha}} \lvaIn{\Psi}_{\alpha}(t, \myvec{x} + \myvec{x}') \myvec{n}_{V_{\alpha}}(t, \myvec{x} + \myvec{x}') \mathrm{d}^{2}\myvec{x}' 
	\nonumber\\
	&\hspace{0.2cm} \approx \int\limits_{A_{\alpha}} \lvaIn{\Psi}_{\alpha}(t, \myvec{x}) \myvec{n}_{V_{\alpha}}(t, \myvec{x} + \myvec{x}') \mathrm{d}^{2}\myvec{x}'
	\nonumber\\
	&\hspace{0.5cm} + \int\limits_{A_{\alpha}} \left( \myvec{x}' \cdot \bn_{\myvec{x}} \right) \lvaIn{\Psi}_{\alpha}(t, \myvec{x}) \myvec{n}_{V_{\alpha}}(t, \myvec{x} + \myvec{x}') \mathrm{d}^{2}\myvec{x}'  
	\nonumber\\
	&\hspace{0.5cm} + \int\limits_{A_{\alpha}} \left[ \frac{1}{2} \left( \myvec{x}' \otimes \myvec{x}' : \bn_{\myvec{x}} \otimes \bn_{\myvec{x}} \right) \lvaIn{\Psi}_{\alpha}(t, \myvec{x}) \cdot \myvec{n}_{V_{\alpha}}(t, \myvec{x} + \myvec{x}') \right] \mathrm{d}^{2}\myvec{x}' 
	\nonumber\\
	&\hspace{0.5cm}+ \dots  \nonumber\\
	~
	&\hspace{0.3cm} \approx \left( \int\limits_{A_{\alpha}} \myvec{n}_{V_{\alpha}} \mathrm{d}^{2}\myvec{x}' \right) \lvaIn{\Psi}_{\alpha} + \left[ \left( \int\limits_{A_{\alpha}} \myvec{x}' \myvec{n}_{V_{\alpha}} \mathrm{d}^{2}\myvec{x}' \right) \cdot \bn_{\myvec{x}} \right] \lvaIn{\Psi}_{\alpha}
	\nonumber \\
	&\hspace{0.2cm} + \frac{1}{2} \left[ \left( \int\limits_{A_{\alpha}} \myvec{x}' \otimes \myvec{x}' \myvec{n}_{V_{\alpha}} \mathrm{d}^{2}\myvec{x}' \right) : \bn_{\myvec{x}} \otimes \bn_{\myvec{x}} \right] \lvaIn{\Psi}_{\alpha}  \, .
	\label{eq:lva:taylor_psi_int1}
\end{align}
We neglect all derivatives beyond second order \cite{Whitaker1999}. The zeroth, first, and second spatial moments of \cref{eq:lva:taylor_psi_int1} are transformed into \cite{Whitaker1999,Quintard1994},
\begin{align}
	&\frac{1}{\abs{V}} \int\limits_{A_{\alpha}} \myvec{n}_{V_{\alpha}} \mathrm{d}^{2}\myvec{x}' = - \bn_{\myvec{x}} \epsA  \, ,  \label{eq:lva:zeroth_intr_moment} \\
	~
	&\frac{1}{\abs{V}} \int\limits_{A_{\alpha}} \myvec{x}' \myvec{n}_{V_{\alpha}} \mathrm{d}^{2}\myvec{x}' = - \bn_{\myvec{x}} \otimes \lvaSup{\myvec{x}_{\alpha}}  \, ,  \label{eq:lva:first_intr_moment} \\
	~
	&\frac{1}{\abs{V}} \int\limits_{A_{\alpha}} \myvec{x}' \otimes \myvec{x}' \myvec{n}_{V_{\alpha}} \mathrm{d}^{2}\myvec{x}' = - \bn_{\myvec{x}} \otimes \lvaSup{\myvec{x}_{\alpha} \otimes \myvec{x}_{\alpha}}  \ .  \label{eq:lva:second_intr_moment}
\end{align}
Inserting these results into \cref{eq:lva:taylor_psi_int1} and dividing by $\abs{V}$ gives
\begin{align}
	&\frac{1}{\abs{V}}  \int\limits_{A_{\alpha}} \lvaIn{\Psi}_{\alpha}(t, \myvec{x} + \myvec{x}') \myvec{n}_{V_{\alpha}}(t, \myvec{x} + \myvec{x}') \mathrm{d}^{2}\myvec{x}' \approx  \nonumber\\
	&\hspace{0.5cm}- \left( \bn_{\myvec{x}} \epsVF_{\alpha} \right) \lvaIn{\Psi}_{\alpha} - \left[ \left( \bn_{\myvec{x}} \otimes \lvaSup{\myvec{x}_{\alpha}} \right) : \bn_{\myvec{x}} \right] \lvaIn{\Psi}_{\alpha}  \nonumber \\
	&\hspace{0.5cm} - \frac{1}{2} \left[ \left( \bn_{\myvec{x}} \otimes \lvaSup{\myvec{x}_{\alpha} \otimes \myvec{x}_{\alpha}} \right) : \bn_{\myvec{x}} \otimes \bn_{\myvec{x}} \right] \lvaIn{\Psi}_{\alpha}  \, .
	\label{eq:lva:taylor_psi_int}
\end{align}
For reasons of simplicity, we drop the function arguments $t$, $\myvec{x}$ and $\myvec{x}'$ in the following sections.

\subsection{Volume-Average of Temporal Evolution Equations}   \label{sec:lva:ex}
In this subsection, we volume-average the various transport and reaction equations occurring in a dynamic battery cell model (see \cref{sec:suppl:soe}).

\subsubsection{Evolution of Volume Fractions}  \label{sec:lva:ex:vf}
In conversion-type batteries, phase volume fractions change in time \cite{Neidhardt2012}. We obtain their temporal evolution by applying the local volume-averaging theory on unity. Setting ${\Psi_{\alpha} = 1}$ in \cref {eq:lva:lva_short,eq:lva:lva_divergence_vector_short,eq:lva:lva_time_derivative_short} yields
\begin{align}
	&\lvaSup{1_{\alpha}} = \frac{1}{\abs{V}} \int\limits_{V_{\alpha}} \mathrm{d}^{3}\myvec{x}' = \frac{V_{\alpha}}{\abs{V}} = \epsA  \, , \label{eq:lva:ex:vf:volume_fraction1}\\
	~
	&\bn \epsA = - \frac{1}{\abs{V}} \int\limits_{A_{\alpha}} \myvec{n}_{V_{\alpha}} \mathrm{d}^{2}\myvec{x}'  \, , \label{eq:lva:ex:vf:nabla_volume_fraction1}\\
	~
	&\partial_{t} \epsA = \frac{1}{\abs{V}} \int\limits_{ A_{\alpha} \backslash \partial\mathcal{D} \hspace*{-5mm} } \left( \velSurf{tot}{\alpha, \alpha'} \cdot \myvec{n}_{\alpha, \alpha'} \right) \mathrm{d}^{2}\myvec{x}'  \, .  \label{eq:lva:ex:vf:time_deriv_volume_fraction1}
\end{align}
We divide the total phase field velocity into its components $\velSurf{$\ast$}{\alpha, \alpha'}$ and $\velSurf{c}{\alpha, \alpha'}$. Because the convective velocity $\velSurf{c}{\alpha, \alpha'}$ is zero on $\partial\mathcal{D}$, we transform the surface integral over $\velSurf{c}{\alpha, \alpha'}$ using the Gaussian theorem (see \cref{eq:lva:ex:vf:time_deriv_volume_fraction1}) into 
\begin{align}
	&\frac{1}{\abs{V}} \int\limits_{ A_{\alpha} \backslash \partial\mathcal{D} \hspace*{-5mm} } \left( \velSurf{c}{\alpha, \alpha'} \cdot \myvec{n}_{\alpha, \alpha'} \right) \mathrm{d}^{2}\myvec{x}'  = \frac{1}{\abs{V}} \int\limits_{ V_{\alpha} } \bn_{\myvec{x}'} \cdot \velSurf{c}{\alpha, \alpha'}(\myvec{x} - \myvec{x}') \mathrm{d}^{3}\myvec{x}'   \, .  \label{eq:lva:ex:vf:time_deriv_volume_fraction2}
\end{align}
Analogous to \cref{eq:lva:transformGradWeightFunc}, we change the differentiation variable and use \cref{eq:lva:intrinsic_lva_short} to obtain
\begin{align}
	&\frac{1}{\abs{V}} \int\limits_{ V_{\alpha} } \bn_{\myvec{x}'} \cdot \velSurf{c}{\alpha, \alpha'}(\myvec{x} - \myvec{x}') \mathrm{d}^{3}\myvec{x}' = -\bn_{\myvec{x}} \left( \frac{\abs{V_{\alpha}}}{\abs{V}} \lvaIn{\velSurf{}{}}^{\mathrm{c}}_{\alpha} \right) = -\bn_{\myvec{x}} \lvaSup{\velSurf{c}{\alpha}}    \, ,  \label{eq:lva:ex:vf:time_deriv_volume_fraction3}
\end{align}
where $\lvaIn{\velSurf{}{}}^{\mathrm{c}}_{\alpha}$ and $\lvaSup{\velSurf{c}{\alpha}}$ denote the intrinsic and superficial volume-average of $\velSurf{c}{\alpha,\alpha'}$ over the phase $\alpha$. Inserting this result in \cref{eq:lva:ex:vf:time_deriv_volume_fraction1} yields
\begin{align}
\partial_{t} \epsilon_{\alpha} &= -\bn_{\myvec{x}} \lvaSup{\velSurf{c}{\alpha}} + \frac{1}{\abs{V}} \int\limits_{A_{\alpha} \backslash \partial \mathcal{D} \hspace*{-5mm}} \left( \velSurf{$\ast$}{\alpha, \alpha'}  \cdot \myvec{n}_{\alpha, \alpha'} \right) \text{d}^{2}\myvec{x}'   \, . \label{eq:lva:ex:vf:time_deriv_volume_fraction}
\end{align}
It is worth mentioning here that $\velSurf{c}{\alpha}$ must not be confused with $\vel_{\alpha}$. $\velSurf{c}{\alpha}$ relates to the phase field $\chi_{\alpha}$, whereas $\vel_{\alpha}$ is the convection velocity of internal quantities, such as concentrations. Nevertheless, in certain cases one can use ${\lvaSup{\velSurf{c}{\alpha}} \approx \lvaSup{\vel_{\alpha}}}$ as an approximation. This would be the case, for example, for rigid bodies.

\subsubsection{Evolution of Concentrations} \label{sec:lva:ex:ce}

Concentrations $c_{\bl}$ in the liquid electrolyte are time evolved with \cite{Schmitt2019}
\begin{align}
	\partial_{t} c_{\bl} + \bn \ncFlux_{\bl} + \bn \left( c_{\bl} \vel_{\liq}  \right) = \source_{\bl}^{\Omega} \, . \label{eq:lva:ex:ce:transport_eq}
\end{align}
The index $\bl$ denotes a solute in the liquid phase. The non-convective fluxes $\ncFlux_{\bl}$ contain diffusion and migration terms. $\vel_{\liq}$ is the convection velocity of the fluid, and $\source_{\bl}^{\Omega}$ is the source term defined inside the volume $\Omega$. Reactions on phase boundaries are not included. We now average \cref{eq:lva:ex:ce:transport_eq} using \cref{eq:lva:lva_short,eq:lva:lva_divergence_vector_short,eq:lva:lva_time_derivative_short}, which results in
\begin{align}
	&\lvaSup{\source_{\bl}^{\Omega}} = \partial_{t} \lvaSup{c_{\bl}} + \bn \lvaSup{\ncFlux_{\bl}} + \bn \lvaSup{ c_{\bl} \vel_{\liq} }   \nonumber\\
	&\hspace{0.5cm} + \frac{1}{\abs{V}} \int\limits_{ A_{\liq} } \left( \ncFlux_{\bl} + c_{\bl} \vel_{\liq} - c_{\bl} \velSurf{c}{\liq} - c_{\bl} \velSurf{$\ast$}{\liq, \alpha'} \right) \myvec{n}_{\liq, \alpha'} \mathrm{d}^{2}\myvec{x}' \, . \label{eq:lva:ex:ce:lva_transport_eq1}
\end{align}
When using \cref{eq:lva:lva_time_derivative_short}, we integrate over the whole surface $A_{\liq}$, since the phase velocity $\velSurf{tot}{\liq, \alpha'}$ is zero on the domain boundary $\partial\mathcal{D}$. To rewrite the integrand of \cref{eq:lva:ex:ce:lva_transport_eq1}, we use the definition of the non-convective flux
\begin{align}
\ncFlux_{\bl} = c_{\bl} (\particleVel_{\bl} - \vel_{\liq})   \, ,
\label{eq:lva:ex:ce:nonConvectiveFlux}
\end{align}
with the particle velocity $\particleVel_{\bl}$. With \cref{eq:lva:ex:ce:nonConvectiveFlux} the integrand of \cref{eq:lva:ex:ce:lva_transport_eq1} transforms to 
\begin{align}
	&\ncFlux_{\bl} + c_{\bl} \vel_{\liq} - c_{\bl} \velSurf{c}{\liq, \alpha'} - c_{\bl} \velSurf{$\ast$}{\liq, \alpha'} =   c_{\bl} \particleVel_{\bl} - c_{\bl} \velSurf{c}{\liq, \alpha'} - c_{\bl} \velSurf{$\ast$}{\liq, \alpha'} \, .
	\label{eq:lva:ex:ce:integrand}
\end{align}
As this flux is evaluated on a phase boundary, we identify it with the flux through the surface. This flux is given by jump mass balance equations \cite{DeVidts1997}. We describe zinc-air batteries in this application, in which each species $\bl$ exists only in a single phase \cite{Schmitt2019,Stamm2017}. Therefore, we can write \cref{eq:lva:ex:ce:integrand} as
\begin{align}
	&\left( c_{\bl} \particleVel_{\bl} - c_{\bl} \velSurf{c}{\liq, \alpha'} - c_{\bl} \velSurf{$\ast$}{\liq, \alpha'} \right) \myvec{n}_{\liq, \alpha'} = -\mathcal{F}_{\bl}^{\liq, \alpha'}  \, ,
	\label{eq:lva:ex:ce:integrand_new}
\end{align}
with the local surface production rate $\mathcal{F}_{\bl}^{\liq, \alpha'}$. Using \cref{eq:lva:ex:ce:integrand_new}, the surface integral in \cref{eq:lva:ex:ce:lva_transport_eq1} becomes
\begin{align}
	&\frac{1}{\abs{V}} \int\limits_{ A_{\liq} } \left( \ncFlux_{\bl} + c_{\bl} \vel_{\liq} - c_{\bl} \velSurf{c}{\liq, \alpha'} - c_{\bl} \velSurf{$\ast$}{\liq, \alpha'} \right) \myvec{n}_{\liq, \alpha'} \mathrm{d}^{2}\myvec{x}'   \nonumber\\
	~
	&=\frac{1}{\abs{V}} \int\limits_{ A_{\liq} } - \mathcal{F}_{\bl}^{\liq, \alpha'} \mathrm{d}^{2}\myvec{x}' = - \frac{|A_{\liq}|}{\abs{V}} \lvaSurfC{\mathcal{F}}_{\bl}^{\liq} = - \lvaSurfC{\source}_{\bl}  \, ,
	\label{eq:lva:ex:ce:surface_integral1}
\end{align}
with the local surface integral $\lvaSurfC{\mathcal{F}}_{\bl}^{\liq}$ (see \cref{eq:lva:lsa_whole_short}) the mean specific surface area of the liquid phase $\abs{A_{\liq}}/\abs{V}$, and the surface source term $\lvaSurfC{\source}_{\bl}$. Since the surface structure is not known exactly, we approximate the local surface integral as a sum over the single surface segments
\begin{align}
	\lvaSurfC{\mathcal{F}}_{\bl}^{\liq} \approx  \frac{1}{|A_{\liq}|} \sum\limits_{\alpha'} \abs{A_{\liq, \alpha'}} \lvaSurfP{\mathcal{F}}_{\bl}^{\liq, \alpha'}  \, ,
	\label{eq:lva:ex:ce:surface_integral2}
\end{align}
where $\abs{A_{\liq, \alpha'}}$ is the surface area between the liquid phase $\liq$ and phase $\alpha'$. Hereby, we define ${\abs{A_{\alpha, \alpha}} = 0}$. Inserting \cref{eq:lva:ex:ce:surface_integral1} in \cref{eq:lva:ex:ce:lva_transport_eq1} results in
\begin{align}
	\partial_{t} \lvaSup{c_{\bl}} + \bn \lvaSup{\ncFlux_{\bl}} + \bn \lvaSup{ c_{\bl} \vel_{\liq} } = \lvaSup{\source_{\bl}^{\Omega}} + \lvaSurfC{\source}_{\bl} \, . \label{eq:lva:ex:ce:lva_transport_eq}
\end{align}
This equation has two types of source terms. $\lvaSup{\source_{\bl}^{\Omega}}$ is related to the bulk volume and $\lvaSurfC{\source}_{\bl}$ is restricted to the interface between two phases. We model the averaged flux and source terms in \cref{sec:lva:mod}.

\subsubsection{Evolution of Electric Potential} \label{sec:lva:ex:pot}
The general form of the evolution equation for the electric potential in phase $\alpha$ is
\begin{align}
	\partial_{t} \rho^{\mathrm{q}}_{\alpha} + \bn \ncFlux_{\mathrm{q}}^{\alpha} + \bn \left( \rho_{\mathrm{q}}^{\alpha} \vel_{\alpha}  \right) = \source_{\mathrm{q}}^{\Omega, \alpha} \, , \label{eq:lva:ex:pot:transport_eq}
\end{align}
with the electric charge density $\rho^{\mathrm{q}}_{\alpha}$. Because this equation has the same structure as \cref{eq:lva:ex:ce:transport_eq}, we apply the same volume-averaging procedure as in \cref{sec:lva:ex:ce}. This results in 
\begin{align}
	\partial_{t} \lvaSup{\rho_{\mathrm{q}}^{\alpha}} + \bn \lvaSup{\ncFlux_{\mathrm{q}}^{\alpha}} + \bn \lvaSup{ \rho_{\mathrm{q}}^{\alpha} \vel_{\alpha} } = \lvaSup{\source_{\mathrm{q}}^{\Omega,\alpha}} + \lvaSurfC{\source}_{\mathrm{q}}^{\alpha} \, . \label{eq:lva:ex:pot:lva_transport_eq1}
\end{align}
We calculate the surface charge source term $\lvaSurfC{\source}_{\mathrm{q}}^{\alpha}$ analogous to \cref{eq:lva:ex:ce:integrand_new,eq:lva:ex:ce:surface_integral1}. If the bulk material is electroneutral, the charge density vanishes, $\rho^{\mathrm{q}}_{\alpha} = 0$. Then \cref{eq:lva:ex:pot:lva_transport_eq1} becomes
\begin{align}
	0 = - \bn \lvaSup{\ncFlux_{\mathrm{q}}^{\alpha}} + \lvaSup{\source_{\mathrm{q}}^{\Omega,\alpha}} + \lvaSurfC{\source}_{\mathrm{q}}^{\alpha} \, . \label{eq:lva:ex:pot:lva_transport_eq}
\end{align}

\subsubsection{Evolution of Convection Velocity} \label{sec:lva:ex:gce}
The multi-component incompressibility constraint (MCIC) \cite{Single2019,Hoffmann2018,Horstmann2013,Stamm2017,Single2016,Single2017} of a fluid is
\begin{align}
	0 = - \bn\vel_{\liq} + \sum\limits_{\bl} \upnu_{\bl} \left[ - \bn \ncFlux_{\bl} + \source_{\bl}^{\Omega} \right]  \, , \label{eq:lva:ex:gce:gce}
\end{align}
with the partial molar volume $\pmv_{\bl}$. By volume-averaging this equation we obtain
\begin{align}
	0 = - \lvaSup{\bn \vel} + \sum\limits_{\bl} \left[ - \lvaSup{\upnu_{\bl} \bn \ncFlux_{\bl}} + \lvaSup{\upnu_{\bl} \source_{\bl}^{\Omega}} \right]  \, . \label{eq:lva:ex:gce:lva_mcic1}
\end{align}
The terms $\lvaSup{\upnu_{\bl} \bn \ncFlux_{\bl}}$ and $\lvaSup{\upnu_{\bl} \source_{\bl}^{\Omega}}$ can only be evaluated with further assumptions. To this aim, we separate the flux and source terms from the partial molar volume. The error of this approximation is limited by the Cauchy-Schwarz inequality \cite{Steele2004}. Since the standard deviation of $\pmv_{\bl}$ is small, the error is also small. Thus, we transform \cref{eq:lva:ex:gce:lva_mcic1} into
\begin{align}
	0 &\approx - \lvaSup{\bn \vel} + \sum\limits_{\bl} \lvaIn{\upnu}_{\bl} \left[ - \lvaSup{\bn \ncFlux_{\bl}} + \lvaSup{\source_{\bl}^{\Omega}} \right]  \nonumber\\
	~
	&= - \bn \lvaSup{\vel} - \frac{1}{\abs{V}} \int\limits_{A_{\liq}} \vel_{\liq} \myvec{n}_{\liq, \alpha'} \mathrm{d}^{2}\myvec{x}'  \nonumber\\
	& + \sum\limits_{\bl} \lvaIn{\upnu}_{\bl} \left[ - \bn \lvaSup{\ncFlux_{\bl}} + \lvaSup{\source_{\bl}^{\Omega}} - \frac{1}{\abs{V}} \int\limits_{A_{\liq}} \ncFlux_{\bl} \myvec{n}_{\liq, \alpha'} \mathrm{d}^{2}\myvec{x}' \right]  \, . \label{eq:lva:ex:gce:lva_mcic2}
\end{align}
With \cref{eq:lva:ex:ce:nonConvectiveFlux} we express $\ncFlux_{\bl} \myvec{n}_{\liq, \alpha'}$ as
\begin{align}
	&\ncFlux_{\bl} \myvec{n}_{\liq, \alpha'} = c_{\bl} \left( \particleVel_{\bl}  - \vel_{\liq} \right)  \myvec{n}_{\liq, \alpha'} =  \nonumber\\
	&\hspace{0.5cm} c_{\bl} \left( \particleVel_{\bl} - \velSurf{tot}{\liq, \alpha'} \right) \myvec{n}_{\liq, \alpha'} + c_{\bl} \left( \velSurf{tot}{\liq, \alpha'} - \vel_{\liq} \right) \myvec{n}_{\liq, \alpha'}  \, .
	\label{eq:lva:ex:gce:Xi_mod}
\end{align}
As in \cref{eq:lva:ex:ce:integrand_new}, we write the first term as the surface source term. Using the incompressibility constraint ($1 = \sum_{\bl} \pmv_{\bl} c_{\bl}$), we transform \cref{eq:lva:ex:gce:lva_mcic2} into
\begin{align}
	0 &= - \bn \lvaSup{\vel} - \frac{1}{\abs{V}} \int\limits_{A_{\liq}} \velSurf{tot}{\liq, \alpha'} \myvec{n}_{\liq, \alpha'} \mathrm{d}^{2}\myvec{x}'  \nonumber\\
	&+ \sum\limits_{\bl} \lvaIn{\upnu}_{\bl} \left[ - \bn \lvaSup{\ncFlux_{\bl}} + \lvaSup{\source_{\bl}^{\Omega}} + \frac{1}{\abs{V}} \int\limits_{A_{\liq}} \mathcal{F}_{\bl}^{\liq, \alpha'} \mathrm{d}^{2}\myvec{x}' \right]  \, . 
	\label{eq:lva:ex:gce:lva_mcic3}
\end{align}
The second term of \cref{eq:lva:ex:gce:lva_mcic3} identifies with the temporal change of the liquid volume fraction (see \cref{eq:lva:ex:vf:time_deriv_volume_fraction}). This yields our final form for the local volume-averaged multi-component incompressibility constraint
\begin{align}
	0 &= - \bn \lvaSup{\vel_{\liq}} + \sum\limits_{\bl} \lvaIn{\upnu}_{\bl} \left[ - \bn \lvaSup{\ncFlux_{\bl}} + \lvaSup{\source_{\bl}^{\Omega}} + \lvaSurfC{\source}_{\bl} \right] - \partial_{t} \epsL \, .
	\label{eq:lva:ex:gce:lva_mcic}
\end{align}

\subsection{Modeling of Transport Fluxes and Reaction Sources} \label{sec:lva:mod}
The local volume-averages (LVA) appearing in \cref{eq:lva:ex:ce:lva_transport_eq,eq:lva:ex:pot:lva_transport_eq,eq:lva:ex:gce:lva_mcic} can only be evaluated with further assumptions. However, we can approximate it by mean quantities. The error of this step is limited by the Cauchy-Schwarz inequality \cite{Steele2004}. We divide the integrals maximally into products of intrinsic volume-averages of bulk variables. For those quantities, the LVA is 
\begin{align}
	&\lvaSup{\Psi_{\ba}} = \epsA \lvaIn{\Psi}_{\ba}  \, . \label{eq:lva:mod:lvaVariable}
\end{align}

\subsubsection{Transport Expressions} \label{sec:lva:mod:tt}
Averaging flux expressions is generally very complicated and tedious. We exemplify our procedure for a diffusive flux in a liquid phase
\begin{align}
	\lvaSup{\ncFlux_{\bl}^{\mathrm{diff}}} = \lvaSup{-D_{\bl} \bn c_{\bl}}  \, . 
	\label{eq:lva:mod:tt:diff_flux0}
\end{align}
We assume that the diffusion coefficient $D_{\bl}$ changes slowly throughout the domain $\mathcal{D}$. Thus, we approximate \cref{eq:lva:mod:tt:diff_flux0} by
\begin{align}
	\lvaSup{\ncFlux_{\bl}^{\mathrm{diff}}} &\approx - \lvaIn{D}_{\bl} \lvaSup{\bn c_{\bl}} = - \lvaIn{D}_{\bl} \left( \bn \lvaSup{c_{\bl}} + \frac{1}{|V|} \int\limits_{A_{\liq}} c_{\bl} \myvec{n}_{\liq, \alpha'} \mathrm{d}^{2}\myvec{x}' \right)  \nonumber \\
	~
	&= - \lvaIn{D}_{\bl} \left( \epsL \bn \lvaIn{c}_{\bl} + \lvaIn{c}_{\bl} \bn \epsL + \frac{1}{|V|} \int\limits_{A_{\liq}} c_{\bl} \myvec{n}_{\liq, \alpha'} \mathrm{d}^{2}\myvec{x}' \right)  \, . 
	\label{eq:lva:mod:tt:diff_flux1}
\end{align}
$\lvaIn{D}_{\bl}$ is the mean diffusion coefficient. It is approximated by $\lvaIn{D(c, \phi)} \approx \lvaIn{D}(\lvaIn{c}, \lvaIn{\phi})$. We adopt the spatial decomposition of \cref{eq:lva:spatial_decomp_psi} and insert \cref{eq:lva:taylor_psi_int} in \cref{eq:lva:mod:tt:diff_flux1}.
\begin{align}
	&\lvaSup{\ncFlux_{\bl}^{\mathrm{diff}}} \approx - \lvaIn{D}_{\bl} \left\lbrace \epsL \bn \lvaIn{c}_{\bl} - \left[ \left( \bn \otimes \lvaSup{\myvec{x}_{\liq}} \right) : \bn \right] \lvaIn{c}_{\bl}  \phantom{\int\limits_{A_{\liq}}}\right.  \nonumber\\
	&- \left. \frac{1}{2} \left[ \left( \bn \otimes \lvaSup{\myvec{x}_{\liq} \otimes \myvec{x}_{\liq}} \right) : \bn \otimes \bn \right] \lvaIn{c}_{\bl} + \frac{1}{|V|} \int\limits_{A_{\liq}} \lvaVar{c}_{\bl} \myvec{n}_{\liq \alpha'} \mathrm{d}^{2}\myvec{x}' \right\rbrace 
	\label{eq:lva:mod:tt:diff_flux2}
\end{align}
Here, the non-physical term ${\lvaIn{c}_{\bl} \bn \epsL}$ vanishes when evaluating the surface integral in \cref{eq:lva:mod:tt:diff_flux1}. In the next step, we discuss the first and second order spatial moments in \cref{eq:lva:mod:tt:diff_flux2}. Whitaker elaborates on this topic in Ref. \cite[chap. 1.3]{Whitaker1999} and Ref. \cite{Quintard1994} in detail. In disordered media, these moments can be completely neglected, as long as the averaging volume is much larger than the typical grain size of the averaging phase. In ordered media, however, they do not vanish completely, but they can be estimated to be small. Ref. \cite{Goyeau1997} shows that certain choices of the weighting function $m$ reduce this contribution. Thus, \cref{eq:lva:mod:tt:diff_flux2} simplifies to
\begin{align}
	\lvaSup{\ncFlux_{\bl}^{\mathrm{diff}}} &\approx - \lvaIn{D}_{\bl} \left[ \epsL \bn \lvaIn{c}_{\bl} + \frac{1}{|V|} \int\limits_{A_{\liq}} \lvaVar{c}_{\bl} \myvec{n}_{\liq, \alpha'} \mathrm{d}^{2}\myvec{x}' \right]  \, . 
	\label{eq:lva:mod:tt:diff_flux3}
\end{align}
We transform \cref{eq:lva:mod:tt:diff_flux3} according to Ref. \cite[chap. 1.4]{Whitaker1999} into 
\begin{align}
	\lvaSup{\ncFlux_{\bl}^{\mathrm{diff}}} &\approx - \lvaIn{D}_{\bl} \epsL \mytens{T}^{\mathrm{eff}}_{\liq} \cdot \bn \lvaIn{c}_{\bl}  \, ,
	\label{eq:lva:mod:tt:diff_flux5}
\end{align}
with the effective transport tensor $\mytens{T}^{\mathrm{eff}}_{\liq}$ of rank 2. Models for this tensor were developed among others by J.C. Maxwell, H.L. Weissberg and M. Quintard \cite{Maxwell1881,Weissberg1963,Quintard1993}. They came up with models for different geometries in different porosity regimes. A popular choice is to assume a macroscopic isotropic medium and reduce $\mytens{T}^{\mathrm{eff}}_{\alpha}$ to a scalar tortuosity factor 
\begin{align}
	\mytens{T}^{\mathrm{eff}}_{\alpha} &\approx \frac{1}{\tau_{\alpha}} \myvec{I} = \epsVF^{b-1}_{\alpha} \myvec{I}  \, ,
	\label{eq:lva:mod:tt:isotropic_eff_transport_tensor}
\end{align}
with the Bruggemann coefficient $\bruggemann$.

The calculation of the other transport expressions is analogous to the provided example.

\subsubsection{Reactions Source Terms} \label{sec:lva:mod:st}

The surface source terms appearing in \cref{eq:lva:ex:ce:lva_transport_eq,eq:lva:ex:pot:lva_transport_eq1} are
\begin{align}
&\lvaSurfC{\source}_{\bl} = \frac{1}{|V|} \sum\limits_{\alpha'} \abs{A_{\liq, \alpha'}} \lvaSurfP{\mathcal{F}}_{\bl}^{\liq, \alpha'}  \, , \label{eq:lva:mod:st:surf_source_term_c}\\
~
&\lvaSurfC{\source}_{\mathrm{q}}^{\alpha} = \frac{1}{|V|} \sum\limits_{\alpha'} \abs{A_{\alpha, \alpha'}} \lvaSurfP{\mathcal{F}}_{\mathrm{q}}^{\alpha, \alpha'}  \, .
\label{eq:lva:mod:st:surf_source_term_q}
\end{align}
These equations state that the total reaction rate emerges from the sum over all surface elements. The surface source term per surface element is the reaction rate per surface area times its specific surface area $\abs{A_{\alpha, \alpha'}}/\abs{V}$. The reaction rates $j^{\alpha, \alpha'}_{i}$ of reaction $i$ at the interface between phase $\alpha$ and $\alpha'$ determine the reaction rate per surface area
\begin{align}
	&\lvaSurfP{\mathcal{F}}_{\ba}^{\alpha, \alpha'} = \sum\limits_{i \in \mathcal{I}} \chi_{\ba, i} j_{i}^{\alpha, \alpha'}  \, , &\lvaSurfP{\mathcal{F}}_{\mathrm{q}}^{\alpha, \alpha'} = \sum\limits_{i \in \mathcal{I}} z_{i}^{e^{+}} F j_{i}^{\alpha, \alpha'}  \, ,
	\label{eq:lva:mod:st:tot_surf_reaction_rate_q}
\end{align}
with the set of all reactions $\mathcal{I}$, the stoichiometric coefficients $\chi_{\ba, i}$ of species $\ba$ in reaction $i$, the Faraday constant $F$ and the charge number $z_{i}^{e^{+}}$. The latter one is the number of virtual, positively charged units, released in this phase by reaction $i$. The volume source term $\lvaIn{\source}_{\mathrm{q}}^{\Omega , \alpha}$ is zero because of the conservation of electric charge.

\section{Numerical Analysis}
\label{sec:num}
In Ref. \cite{Schmitt2019}, we develop a multi-dimensional and physics-based model of zinc-air batteries. Multi-dimensional simulations of such conversion-type batteries are made possible by  the general local volume-averaging theory described in \cref{sec:lva}. However, we find that the resulting system of equations (SOE) is strongly coupled and numerically unstable.

We implement the governing equations (see \cref*{sec:suppl:soe} of the Supplementary Information) in our in-house tool BEST, \emph{Battery and Electrochemistry Simulation Tool} \cite{BEST,Latz2010,Latz2015}. BEST uses a finite volume scheme for spatial discretization and an implicit Euler method for time integration \cite{stabImpl}. The implementation of the time integration is very stable for stiff systems of equations. A Newton iteration produces a linearized system of equations. The solver for the linear system is a multi-grid BiCGSTAB algorithm with an incomplete LU decomposition as preconditioner \cite{Saad,Quarteroni1994,Quarteroni2000,Quarteroni2012,LeVeque2007} and PARDISO \cite{Schenk2004,PARDISO} as coarsest-level solver.

To test the stability of our algorithms we use the setup described in \cref*{sec:suppl:soe} and evaluate different rectilinear geometrical grids (see \cref*{fig:suppl:dd:geomNumStabAna}). We vary the number of voxels ${N_{\mathrm{Y}}}$ and their size ${\Delta \mathrm{Y}}$ in y-direction parallel to the electrodes while keeping the discretization in the orthogonal directions constant. As measure of stability, we choose the maximum time step $\Delta t_{\mathrm{max}}$, for which the solver converges towards a stable solution. If $\Delta t_{\mathrm{max}}$ is greater, the algorithm is more robust. We demand that $\Delta t_{\mathrm{max}}$ exceeds $\SI{e3}{\second}$ because simulations should finish within reasonable time.

\subsection{Initialization}

The initial values of bulk concentrations and volume fractions are not restricted by our model equations. We start from an initially equilibrated state with no reactions, no current, and no fluid flow. Thus, we set ${c^{0}_{\mathrm{s}, \oh} = \lvaIn{c}_{\oh}}$, ${\pressL^{0} = \lvaIn{p}^{\ast}_{\mathrm{atm}}}$, and ${\pressS^{0} = 0}$. The liquid volume fraction $\epsVF^{0}_{\liq}$ is then determined by the Leverett function \cite{Horstmann2013,Whitaker1986} from \cref*{eq:suppl:soe:levertt_function_liq}. The electric potentials are described by elliptic equations and have to be consistent at all times. We determine the initial potentials by solving these nonlinear ellipitic equations. We find that the solver fails to converge for large ratios ${{N_{\mathrm{Y}}} / {\Delta \mathrm{Y}}}$. This corresponds to the white regions in \cref{fig:num:nc:stabAna_classicalCases,fig:num:nc:stabAna_coupled,fig:num:wa:decoupledAlgorithm}. Note that these failed initializations are confined to regions of minor interest and do not obstruct our analysis.

\subsection{Numerical Challenges of a Multi-Component Incompressibility Constraint} \label{sec:num:nc}

In multi-dimensional simulations, the constraint of incompressible media stiffens the system of equations (SOE), such that common algorithms fail to converge. This is even observed for the standard incompressibility constraint ${\bn \lvaSup{\vel_{\liq}} = - \partial_{t} \epsL}$. The multi-component incompressibility constraint (MCIC) \cite{Horstmann2013,Stamm2017,Single2017,Hoffmann2018}, stemming from
\begin{align}
	\sum\limits_{\bl} \lvaIn{\upnu}_{\bl} \lvaIn{c}_{\bl} = 1  \, , \label{eq:num:nc:mcic_base}
\end{align}
worsens this issue (see \cref{eq:lva:ex:gce:lva_mcic}). Most kind of solvers have problems to solve a SOE with a constraint of the form of \cref{eq:lva:ex:gce:lva_mcic}. Thus, we carry out a numerical stability analysis. We examine three scenarios: ${\lvaSup{\vel_{\liq}} = 0}$, ${\bn \lvaSup{\vel_{\liq}}  = -\partial_{t} \epsL}$, and the MCIC (see \cref{eq:lva:ex:gce:lva_mcic}).

\subsubsection{Stability Analysis}
Strong coupling of physico-chemical processes over various length and time scales can prevent the solver from finding a stable solution. Wong et al. \cite{Wong2019} have recently proposed a new sequential-implicit Newton method for solving such systems. However, they focus on simple systems with few variables. We analyze a physically motivated problem of current interest with 13 unknown fields and propose a different method.

\begin{figure}[!tb]
	\centering
	\includegraphics[width=0.9\columnwidth]{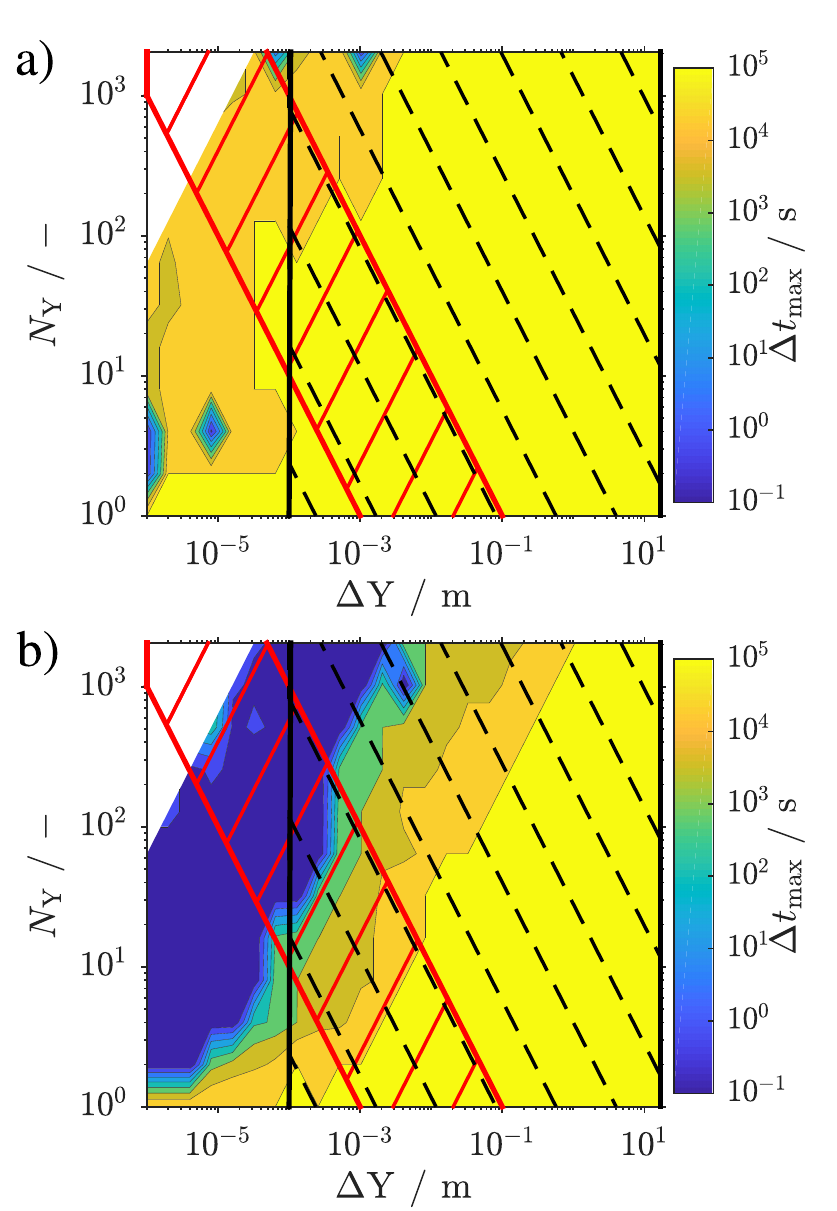}
	\caption{Numerical stability analysis of (a) $\lvaSup{\vel_{\liq}} = 0$ and (b) $\bn\lvaSup{\vel_{\liq}} = -\partial_{t}\epsL$. The maximum possible time step $\Delta t_{\mathrm{max}}$ is a measure of the stability, which is plotted over the number $N_{\mathrm{Y}}$ and size $\Delta \mathrm{Y}$ of the y-direction discretization units. Large values imply a good and small ones a bad stability. Typical configurations ($N_{\mathrm{Y}} \cdot \Delta \mathrm{Y} \approx \SI{0.01}{\meter}$) are within the red shaded region and valid voxel sizes according to the LVA theory ($\Delta \mathrm{Y} \gtrsim \SI{e-4}{\meter}$) in the black shaded region.}
	\label{fig:num:nc:stabAna_classicalCases}
\end{figure}

First, we analyze the simplest case, i.e., ${\lvaSup{\vel_{\liq}} = 0}$. This leads to a simple  Jacobi matrix. We test the stability of the solver on the SOE from \cref*{sec:suppl:soe} by varying geometry and resolution. Figure \ref{fig:num:nc:stabAna_classicalCases}a depicts the result of our stability analysis. For most numerical parameters, our standard solver is very stable. Generally, we find a minor drop of $\Delta t_{\max}$ for large ratios ${{N_{\mathrm{Y}}} / {\Delta \mathrm{Y}}}$. For individual parameter sets, this drop is more significant.

Generally, the best stability is achieved for small $N_{\mathrm{Y}}$ and large ${\Delta \mathrm{Y}}$. Stability worsens for increasing ${{N_{\mathrm{Y}}} / {\Delta \mathrm{Y}}}$. This is because of the higher complexity for a large number of voxels and because of an improper proportion of ${\Delta \mathrm{X}}$, ${\Delta \mathrm{Y}}$, and ${\Delta \mathrm{Z}}$ for a small voxel size in y-direction. However, most of the parameter-space shows a very good stability reaching up to $\SI{e5}{\second}$, which is above the required $\SI{e3}{\second}$. Interesting for our physical system are parameter sets with a valid voxel size according to the local volume-averaging theory ($\Delta \mathrm{Y} \gtrsim \SI{100}{\micro\meter}$) and a reasonable physical extend of the battery ($N_{\mathrm{Y}} \cdot \Delta \mathrm{Y} \approx \SI{1}{\centi\meter}$). These sets lie in the intersection of the black shaded and the red shaded regions in \ref{fig:num:nc:stabAna_classicalCases}. For a vanishing velocity field ${\lvaSup{\vel_{\liq}} = 0}$, this parameter region exhibits sufficient stability.

Second, we test the classical incompressibility-constraint ${\bn \lvaIn{\vel}_{\liq} = -\partial_{t} \epsL}$ and find a reduced stability. The partial time derivative of the electrolyte volume fraction originates here from local volume-averaging of ${\bn \vel_{\liq} = 0}$. We depict the corresponding stability analysis in \cref{fig:num:nc:stabAna_classicalCases}b. A significant drop of stability appears for ${{N_{\mathrm{Y}}} / {\Delta \mathrm{Y}} \gtrsim \SI{e5}{\per \meter}}$. Nevertheless, many of the examined numerical parameter sets and most of the shaded region of interest show a large $\Delta t_{\mathrm{max}}$ and thus a good robustness of the standard algorithm.

\begin{figure}[!tb]
	\centering
	\includegraphics[width=0.9\columnwidth]{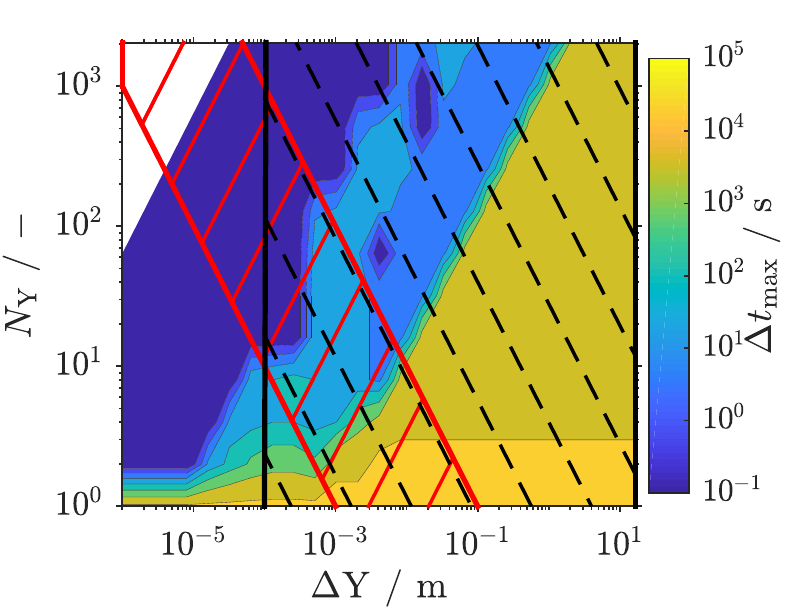}
	\caption{Numerical stability analysis of the MCIC solved by the default algorithm. The maximum possible time step $\Delta t_{\mathrm{max}}$ is a measure of the stability, which is plotted over the number $N_{\mathrm{Y}}$ and size $\Delta \mathrm{Y}$ of the y-direction discretization units. Large values imply a good and small ones a bad stability. Typical configurations ($N_{\mathrm{Y}} \cdot \Delta \mathrm{Y} \approx \SI{0.01}{\meter}$) are within the red shaded region and valid voxel sizes according to the LVA theory ($\Delta \mathrm{Y} \gtrsim \SI{e-4}{\meter}$) in the black shaded region.}
	\label{fig:num:nc:stabAna_coupled}
\end{figure}

Third, we use the MCIC. Figure \ref{fig:num:nc:stabAna_coupled} reveals that the stability of the algorithm drops drastically in this case. An appropriate stability is still achieved for $N_{\mathrm{Y}} = 1$, which corresponds to two-dimensional simulations. This is due to a sparser Jacobian matrix with fewer dependencies. For truly three-dimensional simulations, i.e., $N_{\mathrm{Y}} \geq 2$, the maximum time step ${\Delta t_{\mathrm{max}}}$ decreases strongly for most numerical parameter sets. Most of the physical reasonable configurations are unstable in this case, leading to small time steps of $\Delta t \approx \SI{1}{\second}$. If we use a finer resolution in x- and z-direction, $\Delta t_\mathrm{max}$ is reduced even further. Thus, \cref{fig:num:nc:stabAna_coupled} makes clear that the standard algorithm in BEST is not suited for solving the SOE with sufficiently large time steps.

\subsubsection{Discussion of Stability Analysis}

The observed limited stability has several reasons. First, introducing a pressure field increases the number of unknowns. Second, it increases the number of entries in the Jacobi matrix and makes it denser. Thus, it complicates the linear system of equations, which is passed to the solver. Third and most importantly, the MCIC (see \cref{eq:lva:ex:gce:lva_mcic}) causes a strong coupling of all variables and stiffens the SOE (larger condition number). The last point has the strongest implications. The convection velocity depends on the non-convective fluxes $\lvaSup{\ncFlux_{\bl}}$ and the source terms $\source_{\bl}$ (see \cref{eq:lva:ex:gce:lva_mcic}) in contrast to the standard incompressibility constraint (${\bn\lvaSup{\vel_{\liq}} = -\partial_{t}\epsL}$). Thus, convective and non-convective fluxes are tightly entangled. For example, the local diffusion of a single species causes a change of the liquid volume fraction. This in turn induces a pressure gradient and influences the convection of all species in the whole simulation domain.

The reduction of time steps with increasing resolution is a major challenge. A larger number of grid points implies a longer computation time per iteration. If at the same time $\Delta t_{\mathrm{max}}$ becomes smaller, more iterations have to be carried out to simulate a certain time span. Therefore, the computational effort increases overproportionally with the grid size. In our case, this makes it impossible to perform simulations with a reasonable resolution. Therefore, we develop a novel and robust algorithm for the MCIC in \cref{sec:num:ssi}.

\subsection{Sequential Semi-Implicit Algorithm} \label{sec:num:ssi}

We present in \cref*{sec:suppl:ta} of the supplementary material a set of algorithms, which we tested for the MCIC \cite{Schenk2004,PARDISO,LeVeque2007,Quarteroni2000,Saad,Quarteroni1994,Quarteroni2012,Patankar1980,Ferziger2002,Baumgarte2010}. However, none of them leads to stable simulations. Thus, we develop a new sequential semi-implicit (SSI) algorithm.

We split the system of equations into two subsystems, which are solved separately. The main idea is the separation of the liquid volume fraction and the pressure from the concentrations. This reduces the strong coupling of a fully implicit treatment. The distribution of the unknowns into the two subsystems \textbf{A} and \textbf{B} is shown in \cref{tab:num:wa:subsystems}.
\begin{table}[h!]
	\centering
	\begin{tabular}{l  l}
		\toprule[1.5pt]
		Subsystem \textbf{A}:~~~~~  & Subsystem \textbf{B}:  \\
		$\epsVF_{\zn}$, $\epsVF_{\zno}$, $\epsL$, & $\epsVF_{\zn}$, $\epsVF_{\zno}$, $\lvaSup{\npp}$, \\
		$\pressL$, $\pressS$ & $\lvaIn{c}_{\oh}$, $c_{\mathrm{s}, \oh}$, $\lvaIn{c}_{\znoh}$, $\lvaIn{c}_{\co}$,  $\lvaIn{c}_{\oLiq}$,  \\
		& $\lvaIn{\phi}_{\sol}$, $\lvaIn{\phi}_{\sol}$ 
		\\\bottomrule[1.5pt]
	\end{tabular}	
	\caption[Subsystems of the SSI algorithm.]{Subsystems of the SSI algorithm.}
	\label{tab:num:wa:subsystems}
\end{table}
If solid phase convection is included, it can be advantageous to separate $\pressS$ from $\lvaSup{\npp}$, too. Because of the decoupling of the two subsystems, the multi-component incompressibility constraint is no longer fullfilled numerically. Thus, we introduce the following additional layer of iteration. In each time step, we first solve \textbf{A} partially implicit, with an explicit treatment of the remaining unknowns. Afterwards, we solve \textbf{B} fully implicitly, with updated explicit values from \textbf{A}. Partially implicit means in this context that we use $\epsL$ explicitly in the calculation of the flux expressions. For that we use the previous solution from \textbf{A}. All other occurrences of $\epsL$ and the other variables are implicit. We iterate between the two subsystems, until the solution vector $\unknownVector{}{m}{}$ ($m$: index of SSI iteration) of the whole system of equations converges. We require ${\norm{\unknownVector{}{m+1}{} - \unknownVector{}{m}{}} \leq \epsErr^{\mathrm{dec}} \norm{\unknownVector{}{m+1}{}}}$, with $\epsErr^{\mathrm{dec}} = \SI{1e-4}{}$. If it does not converge within $m^{\mathrm{max}}$ iterations, we reduce the time step $\Delta t$ and try again. If it converges, we increase $\Delta t$ until a maximum predefined time step is reached.

The variables $\epsVF_{\zn}$ and $\epsVF_{\zno}$ loosely couple to the other variables. Thus, we include them in both subsystems. This increases slightly the computational effort, but it represents better the original implicit system of equations and it results in faster convergence.
\begin{figure}[!tb]
	\centering
	\includegraphics[width=0.9\columnwidth]{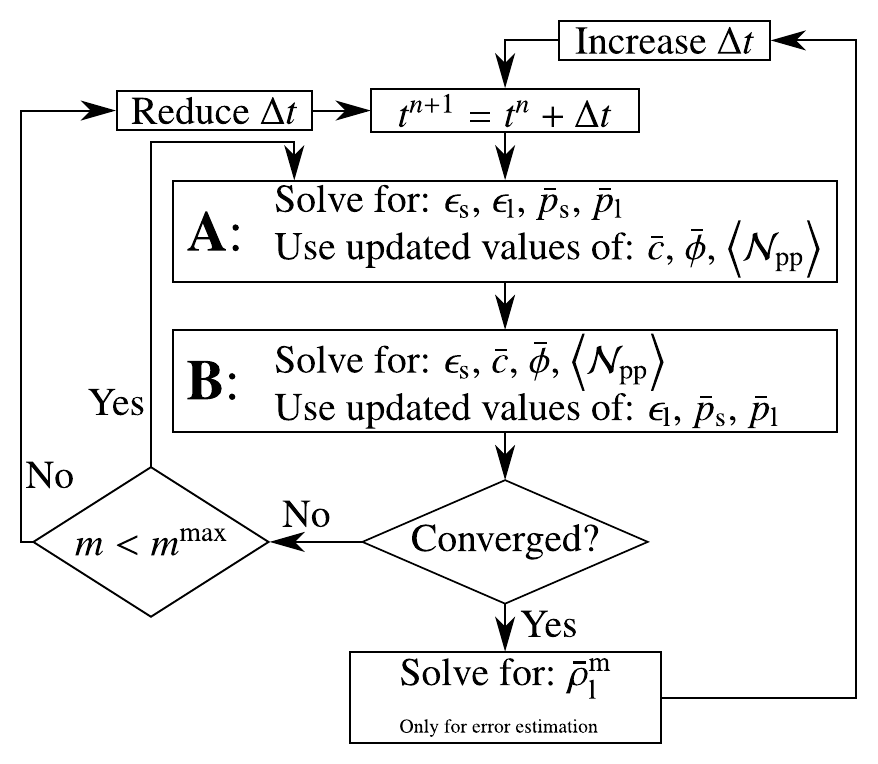}
	\caption{Flowchart of the SSI algorithm. Solving the system of equations partially implicit by two separate subsystems is a stable procedure to cope with a multi-component incompressibility constraint.}
	\label{fig:num:wa:decoupledAlgorithm}
\end{figure}

We show a flowchart of our new SSI algorithm in \cref{fig:num:wa:decoupledAlgorithm}. Our procedure is motivated by the SIMPLE algorithm, which is widely used in computational fluid dynamics \cite{Patankar1980,Ferziger2002,Kajishima2017}. We present the linear and nonlinear solver algorithms in \cref*{sec:suppl:sa}. We include the solid volume fractions in both subsystems in order to speed up the convergence. Additionally, we solve the liquid mass density  $\lvaIn{\rho}^{\mathrm{m}}_{\mathrm{\liq}}$ at the end of each time step by
\begin{align}
	&\partial_{t} \lvaSup{\rho^{\mathrm{m}}_{\liq}} = - \bn \lvaSup{\rho^{\mathrm{m}}_{\liq} \vel_{\liq}} + \sum\limits_{\bl} M_{\bl} \left[ \lvaSup{\source_{\bl}^{\Omega}} + \lvaSurfC{\source}_{\bl} \right] \, , \label{eq:num:wa:equation_rho}
\end{align}
with the updated unknowns of \textbf{A} and \textbf{B}. This step is not necessary for solving the differential equations, because the liquid mass density is determined by the electroneutrality and incompressibility constraints. However, we use this mass density to analyze the numerical accuracy of the SSI algorithm in \cref{sec:num:ea}.

%\subsubsection{Numerical Stability Analysis of the SSI Algorithm}

\begin{figure}[!tb]
	\centering
	\includegraphics[width=0.9\columnwidth]{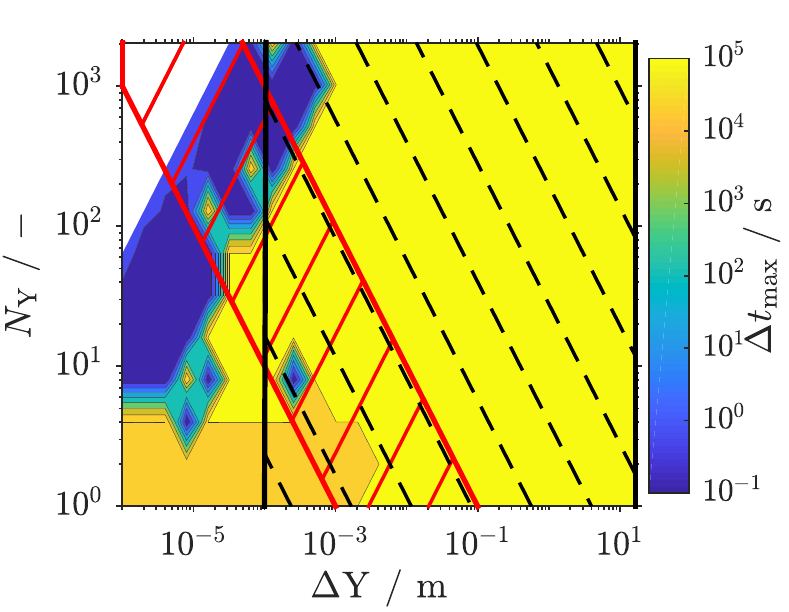}
	\caption{Numerical stability analysis of the MCIC solved with the SSI algorithm. The maximum possible time step $\Delta t_{\mathrm{max}}$ is a measure of the stability, which is plotted over the number $N_{\mathrm{Y}}$ and size $\Delta \mathrm{Y}$ of the y-direction discretization units. Large values imply a good and small ones a bad stability. Typical configurations ($N_{\mathrm{Y}} \cdot \Delta \mathrm{Y} \approx \SI{0.01}{\meter}$) are within the red shaded region and valid voxel sizes according to the LVA theory ($\Delta \mathrm{Y} \gtrsim \SI{e-4}{\meter}$) in the black shaded one.}
	\label{fig:num:wa:stabAna_decoupled}
\end{figure}

Our stability analysis of this newly developed algorithm is depicted in \cref{fig:num:wa:stabAna_decoupled}. We find an improvement by orders of magnitude (in measures of $\Delta t_{\mathrm{max}}$) compared to the fully coupled system in \cref{fig:num:nc:stabAna_coupled}. Most valid parameter sets, determined by the constraints of the local volume-averaging theory, are solved with time steps of $\SI{e5}{\second}$. This is even true for a finer resolution in the x- and z-directions. Due to these improvements, it is now possible to perform multi-dimensional simulations of zinc-air batteries with a multi-component incompressibility constraint \cite{Schmitt2019}.

\subsection{Error Analysis} \label{sec:num:ea}

Decoupling the system of equations introduces an additional numerical error. The multi-component incompressibility constraint \cref{eq:num:nc:mcic_base} is no longer satisfied in each of the subsystems \textbf{A} and \textbf{B}, as liquid volume fraction and pressure are separated from concentrations (see \cref{tab:num:wa:subsystems}). By iterating between the two subsystems, however, we improve accuracy. We check numerically if the solution of our SSI algorithm converges against the solution of the fully implicit algorithm. To this aim, we calculate the liquid mass density $\lvaSup{\rho^{\mathrm{m}}_{\liq}}$ by solving \cref{eq:num:wa:equation_rho} and compare the two independent solutions for the the water concentration, 
\begin{align}
&\lvaIn{c}^{\mathrm{dens}}_{\water} = \frac{1}{M_{\water}} \left( \lvaIn{\rho}^{\mathrm{m}}_{\liq} - \sum\limits_{\bl \neq \water} M_{\bl} \lvaIn{c}_{\bl} \right) \, ,  \label{eq:num:ea:concWaterByDens}\\
~
&\lvaIn{c}^{\mathrm{PMV}}_{\water} = \frac{1}{\lvaIn{\pmv}_{\water}} \left( 1 - \sum\limits_{\bl \neq \water} \lvaIn{\pmv}_{\bl} \lvaIn{c}_{\bl} \right)  \, . \label{eq:num:ea:concWaterByPMV}
\end{align}
If our SSI algorithm generates a consistent solution, as does the fully implicit algorithm, $\lvaIn{c}^{\mathrm{dens}}_{\water}$ and $\lvaIn{c}^{\mathrm{PMV}}_{\water}$ should coincide within the limit of numerical accuracy. We check the consistency of the SSI algorithm for global and local constraints. On the global scale we test if the total liquid volumes, expressed by
\begin{align}
	&V^{\mathrm{dens}}_{\liq, \mathrm{tot}} = \sum\limits^{N}_{i} \sum\limits_{\bl} \epsVF_{\liq, i} \abs{V_{i}} \lvaIn{c}^{\mathrm{dens}}_{\bl, i} \lvaIn{\pmv}_{\bl, i}  \, ,  \label{eq:num:ea:totVolLiqDens}\\
	~
	&V^{\mathrm{PMV}}_{\liq, \mathrm{tot}} = \sum\limits^{N}_{i} \sum\limits_{\bl} \epsVF_{\liq, i} \abs{V_{i}} \lvaIn{c}^{\mathrm{PMV}}_{\bl, i} \lvaIn{\pmv}_{\bl, i}  \,  \label{eq:num:ea:totVolLiqPMV},
\end{align}
are consistent and discuss their deviation
\begin{align}
	E^{\mathrm{glo}} = \frac{V^{\mathrm{dens}}_{\liq, \mathrm{tot}} - V^{\mathrm{PMV}}_{\liq, \mathrm{tot}}}{ \frac{1}{2} \left( V^{\mathrm{dens}}_{\liq, \mathrm{tot}} + V^{\mathrm{PMV}}_{\liq, \mathrm{tot}} \right)}  \,  \label{eq:num:ea:errorMCICglobal}.
\end{align}
The index $i$ runs over all $N$ discretized volume elements with the volumes $\abs{V_{i}}$. However, even a globally consistent solution (${E^{\mathrm{glo}} = 0}$) might violate the MCIC locally. Therefore, we check the local deviation of volumes according to the MCIC (see \cref{eq:num:nc:mcic_base}) by
\begin{align}
	E^{\mathrm{loc}} = \max\limits_{i \in [1, ..., N]} \left( \abs{1 - \sum\limits_{\bl} \lvaIn{\upnu}_{\bl, i} \lvaIn{c}^{\mathrm{dens}}_{\bl, i} } \right) \, .  \label{eq:num:ea:errorMCIClocal}
\end{align}

\begin{figure}[!tb]
	\centering
	\includegraphics[width=\columnwidth]{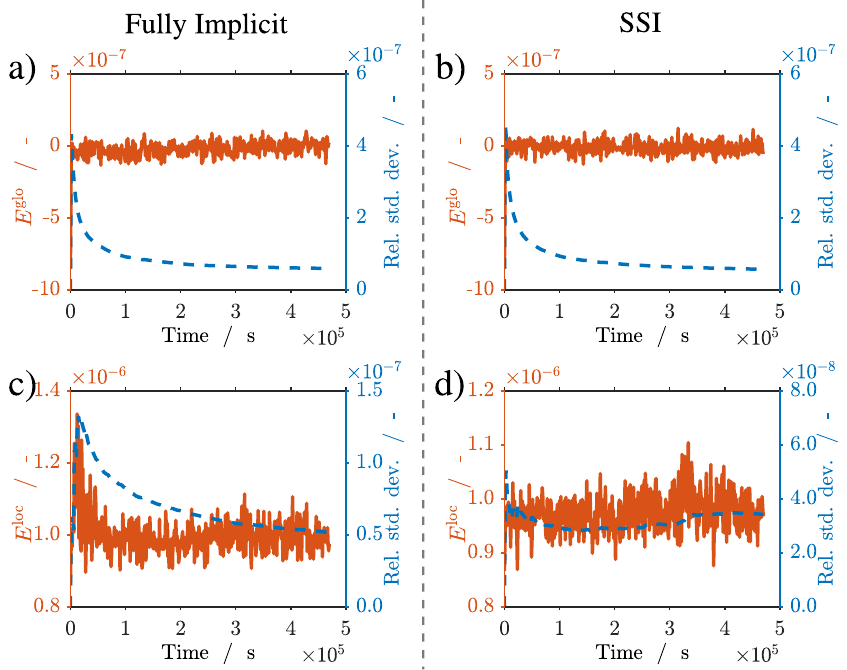}
	\caption{Global and local errors of the fully implicit and the SSI algorithm. (a) and (b) show  the global (see \cref{eq:num:ea:errorMCICglobal}) and (c) and (d) the local (see \cref{eq:num:ea:errorMCIClocal}) error estimates.}
	\label{fig:num:ea:errorGlobalLocal}
\end{figure}

We compare the global (\cref{eq:num:ea:errorMCICglobal}) and the local (\cref{eq:num:ea:errorMCIClocal}) errors of the fully implicit and the SSI algorithms in \cref{fig:num:ea:errorGlobalLocal}. They are on the same orders of magnitude, i.e., $\SI{e-7}{}$ and $\SI{e-6}{}$, for both algorithms. The temporal relative standard deviation of this measure is approximately $\SI{e-7}{}$. To conclude, sequential iteration successfully improves the local and global accuracy to the level of the fully coupled algorithm. 

Furthermore, we monitor the error by checking the total number of hydrogen and zinc atoms, which should be conserved in time. We calculate them with
\begin{align}
	&n_{\hydrogen} = \sum\limits^{N}_{i} \epsVF_{\liq, i} \abs{V_{i}} \left( 2\lvaIn{c}^{\mathrm{PMV}}_{\water, i} + \lvaIn{c}_{\oh, i} + 4\lvaIn{c}_{\znoh, i} \right)  \, , \label{eq:num:ea:aosH}\\
	~
	&n_{\zn} = \sum\limits^{N}_{i} \frac{\epsVF_{\zn, i} \abs{V_{i}}}{\molVol_{\zn}} + \frac{\epsVF_{\zno, i} \abs{V_{i}}}{\molVol_{\zno}} + \epsVF_{\liq, i} \abs{V_{i}} \lvaIn{c}_{\znoh, i}  \, . \label{eq:num:ea:aosZn}
\end{align}
Let ${\tilde{\Psi} = \int^{t_{1}}_{t_{0}} \Psi \mathrm{d}t / (t_{1}-t_{0})}$ be the temporal average of a field variable $\Psi$. The relative errors of the conserved number of hydrogen and zinc atoms are then
\begin{align}
&E^{\hydrogen} = \frac{n_{\hydrogen} - \tilde{n}_{\hydrogen}}{\tilde{n}_{\hydrogen}}     \, ,   \qquad   \qquad          E^{\zn} = \frac{n_{\zn} - \tilde{n}_{\zn}}{\tilde{n}_{\zn}} \, . \label{eq:num:ea:errorAOS}
\end{align}

\begin{figure}[!tb]
	\centering
	\includegraphics[width=\columnwidth]{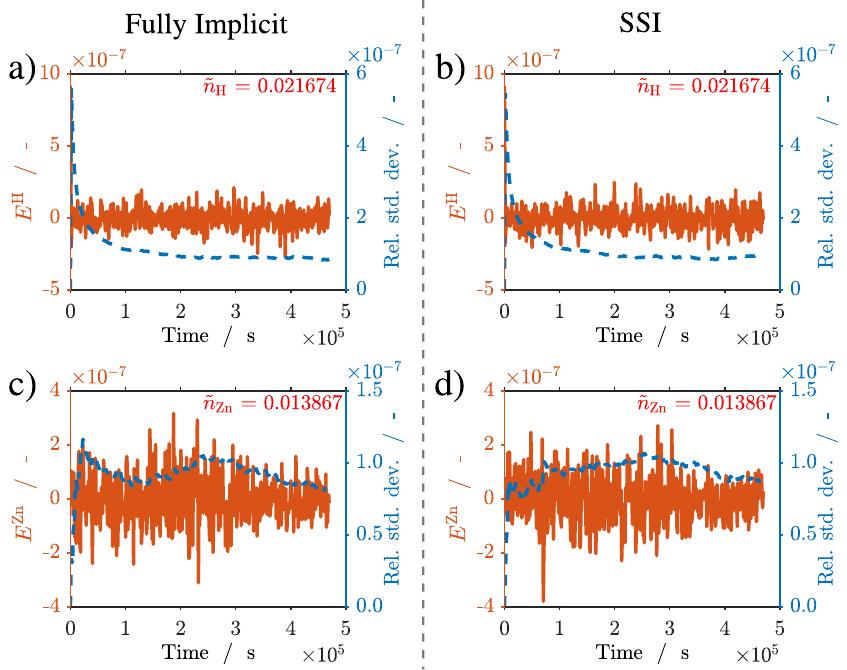}
	\caption{Numerical errors of the physically conserved amount of hydrogen ((a) and (b)) and zinc atoms ((c) and (d)) (\cref{eq:num:ea:errorAOS}) for the fully coupled and the SSI algorithms.}
	\label{fig:num:ea:errorAOS}
\end{figure}

We plot $E^{\hydrogen}$ and $E^{\zn}$ from \cref{eq:num:ea:errorAOS} in \cref{fig:num:ea:errorAOS}. The variations of the particle numbers of hydrogen and zinc are smaller than $\SI{e-7}{}$ with a temporal standard deviation on the order of $\SI{e-7}{}$ for the fully coupled and the SSI algorithms. Thus, our new SSI algorithm conserves the number of particles and its solution satisfies nicely the electroneutrality and incompressibility constraints. It takes on average eleven iterations of the subsystems \textbf{A} and \textbf{B} to achieve this accuracy.

\begin{figure}[!tb]
	\centering
	\includegraphics[width=0.9\columnwidth]{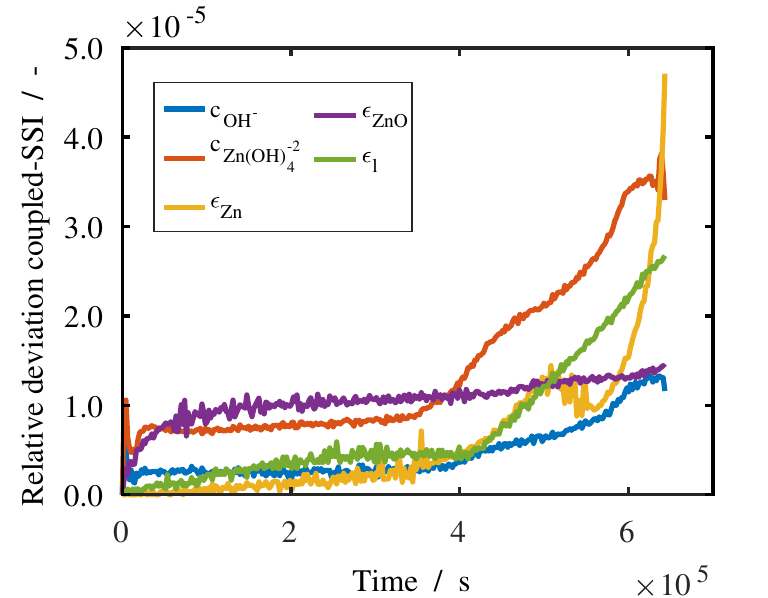}
	\caption[Relative deviation of coupled and decoupled algorithm.]{Maximum relative deviation of SSI and fully coupled algorithms for various independent variables over time.}
	\label{fig:num:ea:convAna}
\end{figure}

Finally, we directly compare the results of the fully coupled and the SSI algorithm. To this aim, we compare the relative deviation for all variables in all discretization units. Figure (\ref{fig:num:ea:convAna}) shows the maximum of this deviation over time. We find that the new algorithm delivers the same result as the fully implicit system with an inaccuracy below \SI{5e-5}{}. The steep increase of the deviation at the end of the simulation is due to the fact that the zinc is nearly completely dissolved and thus $\epsVF_{\zn} \approx 0$.

To conclude, the solution of the SSI algorithm converges against one of the fully implicit algorithm and our new SSI algorithm conserves the physical constraints.

\subsection{Performance Analysis} \label{sec:num:pa}

In the fully coupled algorithm, a single implicit nonlinear system has to be solved per time step. In the SSI algorithm instead, the two nonlinear subsystems \textbf{A} and \textbf{B} are solved iteratively, i.e., multiple times per time step (see \cref{fig:num:wa:decoupledAlgorithm}). This implies a greater computational workload in each time step. However, the systems \textbf{A} and \textbf{B} are smaller and less stiff and convergence is faster. Nevertheless, the SSI system is computationally more expensive per time step. As discussed in \cref{sec:num:nc}, the maximum time step of the coupled algorithm is highly dependent on the grid resolution. Therefore, we examine the dependence of total computation times on grid resolution. 

\begin{figure}[!ht]
	\centering
	\includegraphics[width=0.9\columnwidth]{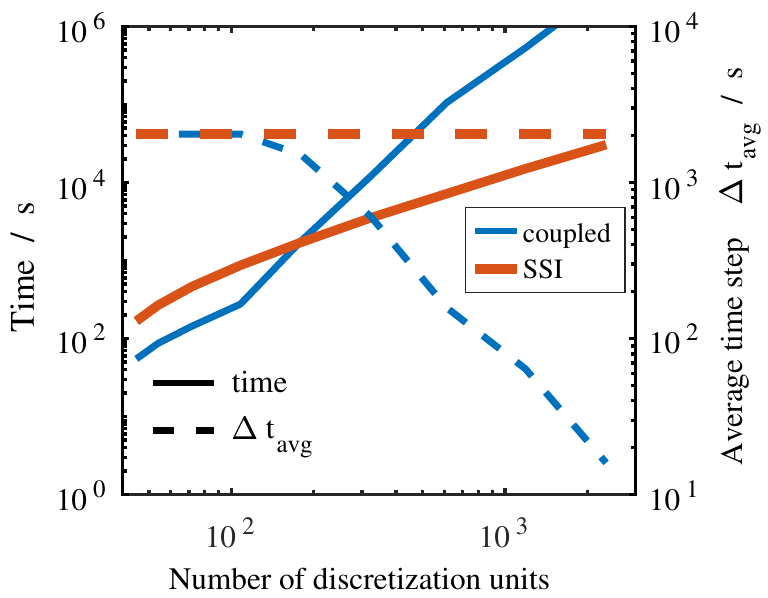}
	\caption{Execution times of the fully coupled and the SSI algorithm for different grid resolutions. The execution time is shown in solid lines and the average time step $\Delta t_{\mathrm{avg}}$ in dashed lines. The maximum possible time step $\Delta t$ is set to \SI{2048}{s}.}
	\label{fig:num:pa:execTimes}
\end{figure}

The results are presented in \cref{fig:num:pa:execTimes}. As test scenario, we choose a 2D simulation of a complete discharge of a zinc-air battery with \SI{710}{\milli\ampere\hour} and a current density of \SI{54.2}{\ampere\per\meter\squared}. We use here a first order Adams-Moulton method for the time integration. We examine the average time step size $\Delta t_{\mathrm{avg}}$ and the total time for completion of the two algorithms, fully coupled and SSI. We find that for less than 100 discretization units, both algorithms reach the maximum specified time step of \SI{2048}{s}. In this case, the SSI algorithm is about three times slower than the fully coupled one. As the number of discretization units increase, the average time step of the fully coupled algorithm decreases linearly and the execution time increases nearly quadratically. In contrast, the average time step of the SSI algorithm remains constant and its total execution time increases only linearly. We conclude that the performance of the SSI algorithm is significantly better than the performance of the fully coupled algorithm for reasonable grid resolutions.

\section{Conclusion}
\label{sec:conc}

In this paper, we have presented a local volume-averaging theory capable of describing time-varying phase compositions. Morphology-changing mechanisms are described by phase convection and heterogeneous reactions. With our newly derived theory, we set up a volume-averaged model of zinc-air conversion batteries. Thus, it is possible to describe and simulate such cells on scales much larger than the pore size.

However, we find that the resulting system of equations is not numerically stable in a multi-dimensional case. The reason for this is the multi-component incompressibility constraint of the concentrated electrolyte. We solve this problem by decoupling the system of equations and solving the subsystems iteratively and semi-implicitly. In a subsequent error analysis, we show that this new algorithm conserves the number of particles, fulfills all constraints, and converges against the fully implicit solution.

\subparagraph*{Conflicts of Interest}~\\

There are no conflicts to declare.

\subparagraph*{Acknowledgments}~\\

The authors thank Simon Hein and Max Schammer for fruitful discussions.
This work was supported by the German Ministry of Education and Research (BMBF) (project LUZI, BMBF: 03SF0499E).
Further support was provided by the bwHPC initiative and the bwHPCC5 project through associated compute services of the JUSTUS HPC facility at the University of Ulm. This work contributes to the research performed at CELEST (Center for Electrochemical Energy Storage Ulm-Karlsruhe).

\subparagraph*{References}~\\

\bibliographystyle{elsarticle-num}
\bibliography{literature}

\end{document}